\documentclass[11pt]{article}
\usepackage{float}
\usepackage[section]{placeins}
\usepackage[utf8]{inputenc}
\usepackage[T1]{fontenc}
\usepackage{lmodern}
\usepackage{amsmath,amssymb,amsfonts}
\usepackage{graphicx}
\usepackage{booktabs}
\usepackage{longtable}
\usepackage{pdflscape}
\usepackage{lscape}
\usepackage{array}
\usepackage{multirow}
\usepackage{authblk}
\usepackage[authoryear,round]{natbib}
\usepackage{geometry}
\usepackage{hyperref}
\usepackage{microtype}
\usepackage{caption}
\usepackage{subcaption}
\usepackage{dblfloatfix}
\usepackage{placeins}
\usepackage{xcolor}

\hypersetup{
  colorlinks=true,
  linkcolor=blue,
  citecolor=blue,
  urlcolor=blue,
  pdftitle={Determination of H0 with a Photometric AGN Golden Sample derived from a joint SN Ia-QSO cosmological reconstruction},
  pdfauthor={S. Rueda-Blanco, M. A. Higuera-G., C. Delgado-Correal}
}

\title{Determination of $H_{0}$ with a Photometric AGN Golden Sample derived from a joint SN Ia-QSO cosmological reconstruction}

\author[1]{S. Rueda-Blanco\thanks{Corresponding author: sruedab@unal.edu.co}}
\author[1]{M. A. Higuera-G.\thanks{Co-author M. A. Higuera-G: mahiguerag@unal.edu.co.}}
\author[2]{C. Delgado-Correal\thanks{Co-author C. Delgado: m.camilo.d@gmail.com}}

\affil[1]{National Astronomical Observatory, National University of Colombia. Bogot\'a, Colombia}
\affil[2]{Faculty of Engineering, Francisco Jos\'e de Caldas District University. Bogot\'a, Colombia.}

\date{Preprint version -- \today}

\begin{document}

\maketitle

\begin{abstract}
The current tension between early and late-Universe determinations of the Hubble constant motivates the development of independent high-redshift distance tracers. Quasars, as luminous active galactic nuclei - AGNs, provide a promising route through the empirical relation between their X-ray and ultraviolet luminosities.We construct a joint Type Ia supernova--quasar Hubble diagram and investigate whether a statistically selected quasar subsample can define a preliminary golden sample for future AGN standardization. We combine calibrated Type Ia supernovae with SDSS DR14 quasars and infer the cosmological parameters from the posterior distribution of a joint model. Posterior medians and credible intervals are obtained from the marginalized distributions. We further define redshift-dependent quasar pools from SDSS queries and select the most representative sources in each bin using a multivariate statistical-distance criterion based on redshift, optical magnitudes, and SDSS colors. The posterior analysis yields
$H_0 = 71.76^{+1.67}_{-1.31}\,\mathrm{km\,s^{-1}\,Mpc^{-1}}$
and
$\Omega_m = 0.294^{+0.014}_{-0.012}$.
The statistically selected quasar candidates reproduce the reference Hubble diagram with small residual dispersion across multiple redshift intervals and remain broadly distributed across the SDSS angular footprint without evidence of strong artificial clustering induced by the selection procedure.
\end{abstract}

\section{Introduction}

The determination of the Hubble constant $H_0$ remains one of the central problems in modern observational cosmology \citep{Planck2020,Riess2022}. Measurements inferred from early-Universe probes, particularly the cosmic microwave background within the standard $\Lambda$CDM framework, yield values near $67$--$68\ \mathrm{km\,s^{-1}\,Mpc^{-1}}$, while late-Universe distance-ladder measurements based on Type Ia supernovae calibrated with Cepheids favor larger values close to $73\ \mathrm{km\,s^{-1}\,Mpc^{-1}}$. This discrepancy, commonly referred to as the Hubble tension, has motivated the search for independent cosmological tracers capable of probing the expansion history across wider redshift intervals \citep{DiValentino2021}.

Quasars constitute one of the most promising alternatives for extending the Hubble diagram beyond the redshift range currently accessible to Type Ia supernovae \citep{Risaliti2015,Lusso2016,Risaliti2019}. As luminous active galactic nuclei (AGN), quasars can be detected up to redshifts $z > 7$, making them potentially powerful probes of the high-redshift Universe \citep{Banados2018}. In recent years, several studies have shown that the empirical non-linear relation between the ultraviolet and X-ray luminosities of quasars may provide a mechanism for constructing standardizable candles and deriving cosmological distance indicators \citep{Lusso2016,Risaliti2019}. In this work we investigate the cosmological applicability of quasars through a joint Type Ia supernova--quasar analysis combining calibrated supernova data with SDSS DR14 quasars. Using posterior inference, we construct a common Hubble diagram and infer the cosmological parameters of the combined model. We additionally introduce a statistical pooling methodology aimed at identifying representative quasar subsamples across different redshift intervals. These statistically selected objects define a preliminary ``golden sample'' intended for future physical validation through spectral energy distribution (SED) fitting and multiwavelength analysis \citep{Ciesla2015,Ramos2020}. The primary objective of this work is not to obtain competitive cosmological constraints, but rather to develop a reproducible framework for identifying statistically representative AGN populations that may serve as future high-redshift cosmological tracers.


\section{Observational data and statistical framework}

\subsection{Cosmological motivation}

The persistent discrepancy between early- and late-Universe
determinations of the Hubble constant has motivated the
development of independent cosmological probes extending
beyond the standard Type Ia supernova framework
\citep{Planck2020,Riess2022,DiValentino2021}. While
supernovae currently provide the most precise late-time
cosmological constraints, their observational reach becomes
progressively limited at high redshift
\citep{Riess2022,Scolnic2022}. In contrast, quasars --- the
luminous manifestation of accreting active galactic nuclei
(AGN) --- can be detected up to redshifts $z>7$, making them
potentially powerful tracers of the expansion history of the
Universe across a much broader cosmological baseline \citep{Banados2018}. Several recent studies have suggested that the empirical
relation between ultraviolet and X-ray luminosities in quasars
may provide a mechanism for constructing standardizable
candles and deriving cosmological distance indicators
\citep{Risaliti2015,Lusso2016,Risaliti2019}. The physical origin of this relation is generally interpreted as a consequence
of the coupling between the thermal emission of the accretion disk and the high-energy emission produced in the hot corona surrounding the central supermassive black hole.
Although the detailed radiative processes remain an active
topic of investigation, the correlation has been observed over
a broad range of luminosities and redshifts, suggesting that it
encodes fundamental properties of the accretion process
\citep{Lusso2016,Nardini2019}. The relation is commonly expressed as

\begin{equation}
\log L_X
=
\gamma \log L_{\mathrm{UV}}
+
\beta ,
\end{equation}

where $L_X$ denotes the monochromatic X-ray luminosity,
$L_{\mathrm{UV}}$ is the ultraviolet luminosity, $\gamma$
represents the slope of the correlation, and $\beta$ is the
intercept. Observational studies typically find values of
$\gamma \simeq 0.6$, indicating a non-linear coupling between
the disk and coronal emission components. Since luminosity and flux are related through \citep{Lusso2016,Risaliti2019}

\begin{equation}
L
=
4\pi d_L^2 F ,
\end{equation}

where $d_L$ is the luminosity distance and $F$ is the observed
flux, the previous relation can be reformulated in terms of
observable quantities and used to estimate cosmological
distances. In practice, the observed ultraviolet and X-ray
fluxes provide indirect information about $d_L$, thereby
allowing quasars to be incorporated into cosmological
analyses over redshift intervals significantly larger than those
currently accessible to Type Ia supernovae. Nevertheless, the intrinsic dispersion associated with quasar
populations remains one of the principal limitations for
precision cosmology \citep{Nardini2019,Bisogni2021}. The
observed scatter reflects the combined influence of
accretion-rate variations, black-hole mass differences,
orientation effects, absorption, and measurement
uncertainties. Consequently, one of the principal challenges
in quasar cosmology is the identification of homogeneous
subsamples whose observational properties remain
sufficiently stable for cosmological applications. We investigate the cosmological applicability of quasars through a joint Type Ia supernova--quasar analysis
combining calibrated supernova data with SDSS DR14 quasars. Using posterior inference, we construct a common cosmological framework in which statistically representative
quasar populations can be identified and evaluated as preliminary high-redshift cosmological tracers.

\subsection{Type Ia supernova cosmological reference}

Type Ia supernovae define the fiducial cosmological reference
adopted throughout the present analysis
\citep{Riess2022,Scolnic2022}. Their calibrated luminosities
allow the construction of precision Hubble diagrams and
provide the observational baseline against which the quasar
candidates are evaluated. Owing to their remarkable
photometric uniformity and the empirical corrections that
standardize their peak luminosities, Type Ia supernovae
currently constitute the most reliable late-time distance
indicators available for cosmological studies. As a result,
they form the foundation of the modern cosmic distance
ladder and provide some of the strongest constraints on the
recent expansion history of the Universe. For each supernova, the observed distance modulus can be
written as

\begin{equation}
\mu
=
m_B
-
M_B ,
\end{equation}

where $m_B$ is the apparent magnitude and $M_B$ is the
absolute magnitude calibration. The corresponding theoretical
distance modulus is given by \citep{Hogg1999}

\begin{equation}
\mu_{\mathrm{th}}(z)
=
5\log_{10}
\left(
\frac{d_L(z)}
{\mathrm{Mpc}}
\right)
+
25 ,
\end{equation}

where $d_L(z)$ is the luminosity distance.

Assuming a spatially flat $\Lambda$CDM cosmology, the
luminosity distance is expressed as

\begin{equation}
d_L(z)
=
\frac{c(1+z)}
{H_0}
\int_0^z
\frac{dz'}
{
\sqrt{
\Omega_m(1+z')^3
+
(1-\Omega_m)
}
},
\end{equation}

where $H_0$ is the Hubble constant, $\Omega_m$ is the
present-day matter density parameter, and $c$ is the speed
of light. The dependence of $d_L$ on these parameters
establishes the direct connection between the observed
supernova magnitudes and the underlying cosmological
model. The comparison between the observed and theoretical
distance moduli constitutes the basis of cosmological parameter estimation. In the present work, the supernova sample provides the low-redshift anchor of the combined
SN Ia--QSO Hubble diagram and defines the fiducial distance--redshift relation used throughout the subsequent quasar analysis. The inferred cosmological reconstruction therefore remains fundamentally tied to the supernova
calibration, while the quasar population serves to extend the observational coverage toward significantly higher redshifts than those densely populated by supernovae alone.

\subsection{SDSS DR14 quasar sample}

The quasar dataset employed in this work was constructed
from the Sloan Digital Sky Survey (SDSS) DR14 quasar
catalog \citep{Paris2018}. Spectroscopic redshifts together
with SDSS optical photometry were extracted through dedicated SQL queries over multiple cosmological intervals. The analysis incorporates the SDSS photometric bands
$u$, $g$, $r$, $i$, and $z$
\citep{Fukugita1996,York2000},
which provide homogeneous optical coverage suitable for statistical population analyses. From these bands, several optical color indices were constructed in order to characterize
the spectral distribution of the quasar population in optical wavelength space \citep{Richards2001}. The adopted color
combinations are $(u-g), (g-r), (r-i), (i-z).$ These colors provide a compact representation of the
observable spectral energy distribution and are sensitive to
physical properties such as accretion state, reddening,
continuum shape, and emission-line contributions. As a
result, color information constitutes a useful parameter space
for identifying statistically representative members of the
parent quasar population. For each object, a multidimensional observable vector $\textbf{x}$ was
defined as

\begin{equation}
\mathbf{x}
=
\left(
z,
u,
g,
r,
i,
z_{\rm SDSS},
u-g,
g-r,
r-i,
i-z_{\rm SDSS}
\right),
\end{equation}

where $z$ denotes the spectroscopic redshift and
$z_{\rm SDSS}$ corresponds to the SDSS $z$-band
magnitude. This vector provides the basis for the multidimensional statistical selection procedure described
below. The parent quasar population was subsequently separated into independent redshift intervals in order to investigate
the stability of the statistical selection procedure across different cosmological epochs. Because quasar populations exhibit significant luminosity and number-density evolution
with redshift \citep{Caplar2015,Caplar2018}, the redshift partitioning was designed to sample both intermediate and high-redshift regimes while preserving sufficient
population statistics within each interval built for this work. This binning strategy allows the statistical properties of the quasar population to be evaluated as a function of cosmic evolution while minimizing biases associated with
strongly heterogeneous redshift distributions.

\subsection{Joint posterior cosmological inference}

The cosmological calibration was performed through a joint Bayesian analysis combining the Type Ia supernova and quasar samples within a common statistical framework (Gelman et al. 2013; Risaliti \& Lusso 2019). The model simultaneously constrains the cosmological parameters $H_0$ and $\Omega_m$, together with the nuisance parameters $M_B$ and $\Delta Q$. Here, $M_B$ corresponds to the absolute-magnitude calibration of the supernova sample, while $\Delta Q$ represents the global offset required to project the quasar population onto the same distance-modulus scale as the supernova sample. In this sense, $\Delta Q$ absorbs the relative calibration difference between the two populations and enables the construction of a common SN Ia--QSO Hubble diagram. The parameter vector is therefore defined as

\begin{equation}
\Theta = (H_0,\Omega_m,M_B,\Delta Q).
\end{equation}

The parameter ($\Delta Q$) represents the global calibration offset required to place the quasar distance indicator on the same distance-modulus scale as the Type Ia supernova sample. Unlike ($M_B$), which is associated with the absolute-magnitude calibration of supernovae, ($\Delta Q$) does not correspond to an intrinsic luminosity of individual quasars. Instead, it absorbs the global normalization of the UV--X-ray luminosity relation, including the intercept ($\beta$), unit conventions, and the relative calibration between the quasar and supernova distance scales. In this sense, ($\Delta Q$) acts as an effective population-level calibration parameter for the quasar Hubble diagram. Assuming independent Gaussian observational uncertainties, the combined likelihood is defined as

\begin{equation}
\mathcal{L}(\Theta)
\propto
\exp\left(-\frac{\chi^2}{2}\right),
\end{equation}

where $\chi^2=\chi^2_{\rm SN}+\chi^2_{\rm QSO}.$ 
The supernova contribution here is

\begin{equation}
\chi^2_{\rm SN}
=
\sum_i
\frac{
\left[
\mu_{{\rm SN},i}
-
\mu_{\rm th}(z_i)
\right]^2
}
{\sigma_{{\rm SN},i}^{\,2}},
\end{equation}

while the quasar contribution is

\begin{equation}
\chi^2_{\rm QSO}
=
\sum_j
\frac{
\left[
\mu_{{\rm QSO},j}
-
\mu_{\rm th}(z_j)
-
\Delta Q
\right]^2
}
{\sigma_{{\rm QSO},j}^{\,2}}.
\end{equation}

The theoretical distance modulus $\mu_{\rm th}(z)$ is computed from the flat $\Lambda$CDM cosmological model described in Sect.~2.2. The combined likelihood therefore simultaneously constrains the cosmological expansion history and the relative normalization between the supernova and quasar populations. The full algebraic development of the luminosity-distance relation, the distance-modulus formalism, the quasar UV--X-ray distance indicator, and the joint SN Ia--QSO likelihood is provided in Appendix~A.

Posterior distributions were obtained through Bayesian sampling of the joint likelihood. The posterior medians together with the 16th and 84th percentiles were adopted as parameter estimators and credible intervals. The resulting posterior distributions define the fiducial cosmological framework used throughout the remainder of this work.

\subsection{Construction of statistical quasar pools}

A central objective of this work is the identification of
statistically representative quasar candidates suitable for
future physical calibration as cosmological tracers. For each
redshift interval, multidimensional statistical pools were
constructed using

\begin{enumerate}
\item spectroscopic redshift,
\item SDSS optical magnitudes,
\item and SDSS optical color indices.
\end{enumerate}

The motivation for this approach arises from the intrinsic
heterogeneity of quasar populations. Quasars span broad
ranges of luminosity, accretion rate, black-hole mass,
orientation, and spectral properties, all of which may
contribute to the observed dispersion in cosmological
applications. Consequently, rather than selecting objects
according to a single observable quantity, we characterize
each quasar through a multidimensional parameter space
containing both photometric and color information. For each redshift interval, the multidimensional observable
vector was defined as follows in the observable vector $\textbf{x}$, supplemented by the spectroscopic redshift of the source.
Median properties and dispersions were computed
independently for each pool in order to characterize the
central behavior of the parent quasar population within the
multidimensional parameter space. A statistical distance metric $D^2$ was then defined as

\begin{equation}
D^2
=
\sum_i
\left(
\frac{x_i-\tilde{x}_i}
{\sigma_i}
\right)^2 ,
\end{equation}

where $x_i$ represents the observed properties of each quasar, $\tilde{x}_i$ denotes the median pool properties, and $\sigma_i$ corresponds to the associated statistical dispersion. This metric measures the deviation of each object relative to the characteristic properties of the parent population. Equation (11) may be interpreted as a simplified form of the Mahalanobis distance in which the covariance structure is approximated by the individual parameter dispersions \citep{Mahalanobis1936}. Consequently, objects with small values of $D^2$ occupy regions of parameter space located near the statistical center of the population, whereas large values identify increasingly atypical or extreme sources. The underlying assumption of the present methodology is that quasars located near the multidimensional center of the parent population are less likely to represent rare subclasses or strongly peculiar objects. Such sources may therefore provide a more stable basis for cosmological applications. By selecting objects whose photometric and color properties remain simultaneously close to the median behavior of the population, the procedure seeks to reduce the impact of intrinsic heterogeneity while preserving the dominant characteristics of the parent sample. A derivation of the multidimensional statistical-distance criterion from the diagonal approximation to the Mahalanobis distance, together with the compact mathematical workflow of the selection procedure, is given in Appendix~A.

Objects with the smallest values of $D^2$ were interpreted as the most statistically representative members of the corresponding parent population. These candidates define the preliminary quasar ``golden sample'' used throughout the present analysis. The statistically selected quasars were subsequently projected onto the combined supernova--quasar Hubble diagram in order to evaluate their cosmological consistency relative to the posterior-calibrated model
\citep{Risaliti2019}. Additional validation was performed through residual distance-modulus analyses and Aitoff sky projections. The sky projections additionally allow verification that the statistical selection does not artificially introduce strong angular clustering within the SDSS footprint. The selected candidates reproduce the fiducial Hubble relation with relatively small residual dispersion across multiple redshift intervals, supporting the applicability of quasars as high-redshift cosmological tracers \citep{Bisogni2021}. The statistical properties of the resulting sample are analyzed in detail in Sect.~4.




\section{Cosmological Calibration from the Joint SN Ia-QSO Analysis}

\subsection{Primary SN Ia and SDSS DR14 Quasar Samples}

The cosmological calibration adopted in this work is first constructed from a primary sample composed of Type Ia supernovae and spectroscopically confirmed SDSS DR14 quasars \citep{Riess2022,Scolnic2022,Paris2018}. The Type Ia supernova sample provides the low-redshift distance scale and defines the calibrated reference relation in the region where supernova cosmology is best constrained \citep{Riess2022,Scolnic2022}. The SDSS DR14 quasar sample extends this relation toward higher redshifts, allowing the Hubble diagram to be explored beyond the range densely populated by supernovae alone \citep{Paris2018,Risaliti2019,Lusso2020}. This primary sample is used only to infer the fiducial cosmological relation. The photometric AGN candidates selected from SDSS DR17 are not included at this stage. This separation is important because the DR14 spectroscopic quasar sample provides the calibrated cosmological baseline, while the DR17 photometric sample is subsequently evaluated against that baseline. In this way, the construction of the golden sample is anchored to an independently calibrated SN Ia--QSO Hubble relation rather than being defined directly from the same photometric candidates used for validation.

The primary cosmological calibration sample consists of the Pantheon+SH0ES Type Ia supernova compilation together with the spectroscopically confirmed SDSS DR14 quasar catalog. The supernova sample provides a densely populated low-redshift distance ladder extending up to $z \sim 2.3$, while the SDSS DR14 quasars extend the observational coverage toward significantly higher redshifts, reaching $z > 5$. The combined dataset therefore spans more than an order of magnitude in cosmological distance and provides the broad redshift baseline required for the joint SN Ia–QSO calibration (Riess et al. 2022; Scolnic et al. 2022; Pâris et al. 2018; Risaliti \& Lusso 2019). 

The joint use of supernovae and quasars also provides the nuisance calibration required to place both populations on a common distance-modulus scale \citep{Lusso2016,Risaliti2019}. The supernova absolute-magnitude parameter $M_B$ fixes the low-redshift calibration, while the quasar offset parameter $\Delta Q$ accounts for the relative projection of the quasar distance indicator onto the supernova-calibrated Hubble diagram \citep{Lusso2016,Risaliti2019}. Once these quantities are inferred, the resulting cosmological relation can be used as the fiducial model for the photometric selection described in Sect.~4. The calibration procedure therefore establishes a cosmological bridge between the low-redshift supernova distance ladder and the high-redshift quasar population. This common framework enables the subsequent evaluation of photometric AGN candidates relative to an independently calibrated Hubble relation, ensuring that the construction of the golden sample remains statistically separated from the cosmological inference itself.

\subsection{Cosmological Constraints from the Joint SN Ia-QSO Sample}

The joint Type Ia supernova--quasar posterior analysis provides simultaneous constraints on the cosmological parameters $H_0$ and $\Omega_m$, together with the nuisance calibration parameters associated with the combined Hubble-diagram reconstruction. The posterior distributions were obtained from the combined likelihood of the supernova and quasar samples under the flat $\Lambda$CDM framework described in Sect.~2 \citep{Planck2020}. Posterior medians together with the 16th and 84th percentiles were adopted as parameter estimators and credible intervals \citep{Gelman2013}. The resulting cosmological constraints indicate that the combined SN Ia--QSO reconstruction remains statistically compatible with late-Universe cosmological measurements while simultaneously extending the Hubble diagram toward significantly higher redshifts \citep{Riess2022,Risaliti2019,Lusso2020}. Table~\ref{tab:cosmo_constraints} summarizes the principal posterior constraints obtained from the joint analysis.

\begin{table}[tbh]
\centering
\caption{Cosmological and calibration parameters inferred from the joint SN Ia--QSO analysis. Reported values correspond to the posterior median together with the 16th and 84th percentiles of the marginalized distributions. These parameters define the fiducial cosmological relation adopted throughout the photometric AGN selection procedure.}
\label{tab:cosmo_constraints}
\begin{tabular}{lcc}
\hline
Parameter & Median & 68\% credible interval \\
\hline
$H_0$ & $71.76$ & $^{+1.67}_{-1.31}$ \\
$\Omega_m$ & $0.294$ & $^{+0.014}_{-0.012}$ \\
$M_B$ & $-19.30$ & $^{+0.05}_{-0.04}$ \\
$\Delta Q$ & $102.47$ & $^{+0.05}_{-0.04}$ \\
\hline
\end{tabular}
\end{table}

The inferred cosmological solution defines the fiducial distance--redshift relation adopted throughout the remainder of this work. Rather than treating the posterior parameters exclusively as cosmological constraints, we use them as a calibrated reference against which the statistical consistency of independent AGN populations can be evaluated. This approach allows the subsequent photometric selection procedure to be anchored to an empirically calibrated Hubble relation derived from the combined SN Ia and spectroscopic quasar dataset.

The inferred value of $H_0$ lies between the values commonly associated with early-Universe CMB measurements and local late-Universe determinations \citep{Planck2020,Riess2022}. Similarly, the inferred matter density parameter remains consistent with the expectations of a standard flat $\Lambda$CDM cosmology \citep{Planck2020}. Although the primary objective of the present work is not to provide competitive cosmological constraints, the resulting parameter values establish a physically plausible and internally consistent framework within which the statistical properties of the photometric AGN candidates can be investigated. In this sense, the inferred cosmological model acts as a bridge between the spectroscopically calibrated sample and the photometric AGN candidates investigated in the following section. The posterior solution therefore provides both the cosmological baseline and the distance--redshift reference relation used to evaluate the compatibility of the selected AGN population with the combined SN Ia--QSO Hubble diagram.

\subsection{Confidence Regions and Parameter Correlations}

The inferred posterior distributions additionally allow the statistical correlations between the cosmological and nuisance parameters to be investigated. In particular, the joint posterior structure reveals the degeneracies associated with the combined calibration of the supernova and quasar populations.

\begin{figure}[tbh]
\centering
\includegraphics[width=0.7\linewidth]{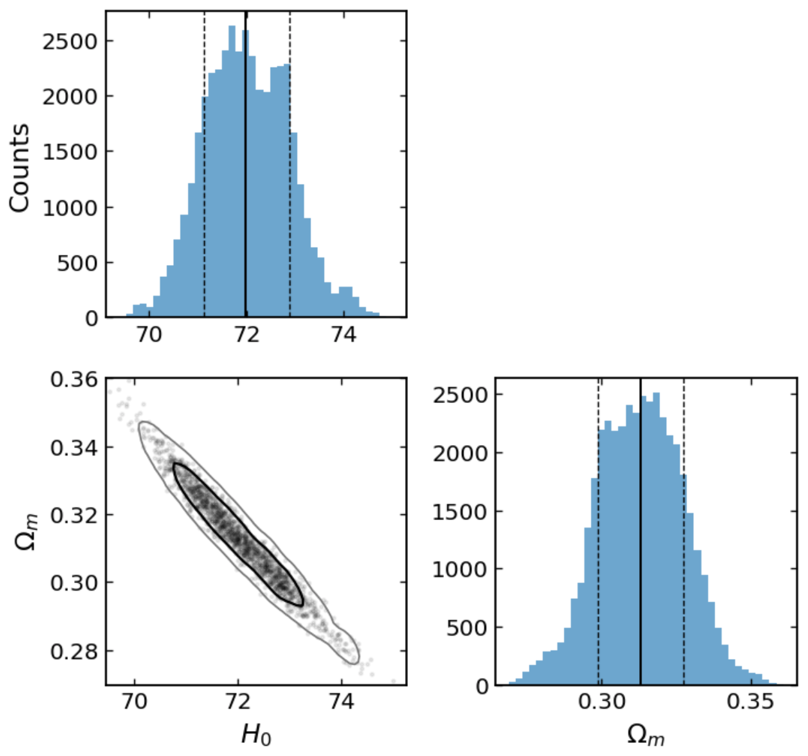}
\caption{Marginalized posterior distributions and confidence regions obtained from the joint SN Ia-QSO cosmological analysis. The contours indicate the 68\% and 95\% credible regions. The inferred cosmological solution defines the fiducial distance-redshift relation used as the reference framework for the photometric AGN selection procedure. $\Omega_m$ denotes the matter density parameter and $H_0$ the Hubble constant.}
\label{fig:corner_joint}
\end{figure}

Figure~\ref{fig:corner_joint} presents the posterior marginalized distributions and confidence contours obtained from the joint cosmological analysis. The posterior distributions indicate that the combined supernova--quasar reconstruction produces a stable cosmological solution within the explored parameter space. In particular, the inferred value of $H_0$ remains intermediate between the values commonly associated with early-Universe CMB measurements and local late-Universe determinations \citep{Planck2020,Riess2022}. The shape of the confidence contours follows the expected degeneracy structure of flat $\Lambda$CDM distance-modulus reconstructions. Variations in $H_0$ can be partially compensated by changes in $\Omega_m$, producing the elongated confidence regions observed in the two-dimensional projections \citep{Planck2020}. Despite these correlations, the posterior distributions remain well localized and exhibit no evidence of disconnected high-probability regions. This behavior indicates that the combined SN Ia--QSO calibration identifies a stable cosmological solution suitable for subsequent photometric validation tests. The posterior distributions additionally illustrate the statistical coupling between the cosmological parameters and the nuisance calibration terms. In particular, the parameters $M_B$ and $\Delta Q$ remain correlated with the cosmological reconstruction because both quantities contribute to the overall normalization of the combined Hubble diagram. The resulting confidence regions therefore provide a direct visualization of the calibration process that links the supernova and quasar populations within a common cosmological framework. The absence of multimodal structures in the posterior distributions further suggests that the combined dataset converges toward a unique and internally consistent solution. Consequently, the inferred cosmological relation can be adopted as a robust fiducial reference for evaluating the statistical compatibility of the photometric AGN candidates to be further investigated. The confidence regions shown also provide quantitative information on the uncertainty associated with the fiducial cosmological reconstruction adopted throughout this work. Rather than serving solely as a cosmological constraint, the posterior solution defines the statistical reference against which the photometric AGN candidates are evaluated in the subsequent selection procedure. Consequently, the confidence contours establish the level of robustness of the distance--redshift relation used to identify statistically compatible photometric quasars. The absence of disconnected high-probability regions further indicates that the inferred cosmological baseline is sufficiently well constrained to support the multidimensional pooling and projection-consistency analyses presented in the following sections.

\subsection{Projection-Consistency Criterion}

The posterior-calibrated cosmological reconstruction obtained from the joint SN Ia--QSO analysis defines the fiducial distance--redshift relation adopted throughout the present work (Risaliti \& Lusso 2019; Bisogni et al. 2021). Once the photometric candidate pool has been identified through the multidimensional statistical-distance analysis described in Sect.~2.5, the cosmological consistency of the selected objects can be evaluated relative to this reference model. For each photometric candidate, the projected distance modulus was obtained by interpolating the posterior-calibrated cosmological reconstruction at the corresponding redshift. The consistency of each candidate with the fiducial cosmological relation was then quantified through the residual distance modulus

\begin{equation}
\Delta\mu
=
\mu_{\mathrm{candidate}}
-
\mu_{\mathrm{posterior}},
\end{equation}

where $\mu_{\mathrm{posterior}}$ denotes the distance modulus predicted by the cosmological reconstruction and $\mu_{\mathrm{candidate}}$ represents the projected distance modulus associated with the photometric candidate. Objects exhibiting small values of $|\Delta\mu|$ are therefore statistically more consistent with the calibrated cosmological reconstruction (Lusso \& Risaliti 2016; Risaliti \& Lusso 2019). In order to evaluate the robustness of the selection procedure, several residual-consistency windows were explored. For each adopted threshold $|\Delta\mu|_{\max}$, the number of surviving candidates and the corresponding residual statistics were computed.

\begin{table}[tbh]
\centering
\caption{Sensitivity of the photometric AGN sample to the adopted projection-consistency window. The number of selected candidates increases with the residual threshold. A fiducial value of $|\Delta\mu|\leq 0.02$ mag was adopted throughout this work.}
\label{tab:projection_sensitivity}
\begin{tabular}{ccccc}
\hline
$|\Delta\mu|_{\rm max}$ (mag) &
$N_{\rm selected}$ &
$\langle \Delta\mu \rangle$ (mag) &
$\sigma_{\Delta\mu}$ (mag) & $RMS_{\Delta\mu}$\\
\hline
0.005 & 71 & 0.00016 & 0.00313 & 0.00311 \\
0.010 & 138 & 0.00031 & 0.00575 & 0.00574 \\
0.020 & 255 & 0.00033 & 0.01094 & 0.01093 \\
0.050 & 346 & -0.00004 & 0.01800 & 0.01797 \\
0.100 & 350 & 0.00000 & 0.01912 & 0.01909 \\
\hline
\end{tabular}
\end{table}

Table~\ref{tab:projection_sensitivity} shows that the number of selected candidates increases monotonically as the residual-consistency window becomes less restrictive. At the same time, the residual dispersion increases as progressively larger deviations from the fiducial cosmological relation are allowed. This behavior is expected and reflects the trade-off between sample size and cosmological consistency (Gelman et al. 2013). A fiducial threshold of $|\Delta\mu| \leq 0.02$ mag was adopted throughout the remainder of this work. This value provides a compromise between retaining a statistically significant number of objects and preserving a relatively small residual dispersion with respect to the posterior-calibrated Hubble relation. The resulting sample defines the photometric AGN golden sample investigated in Sect.~4. Similar approaches based on residual analyses and dispersion minimization have been shown to be effective for identifying statistically coherent quasar populations suitable for cosmological applications (Nardini et al. 2019; Bisogni et al. 2021). It is important to emphasize that the projection-consistency criterion is not intended as an independent cosmological validation of the selected AGN population. Rather, it acts as a filtering condition designed to identify candidates whose projected distance moduli remain statistically compatible with the fiducial SN Ia–QSO reconstruction. Consequently, the subsequent agreement between the selected candidates and the calibrated Hubble relation should be interpreted as a consistency requirement imposed by the selection procedure itself.

\section{Cosmological Selection of a Photometric AGN Golden Sample}

\subsection{Redshift Distribution and Sample Composition}

The stability of the multidimensional pooling framework and the subsequent cosmological consistency selection were additionally investigated through the relative distribution of selected candidates across the explored redshift intervals. Because quasar populations evolve significantly with redshift, it is necessary to verify that the statistical selection procedure remains sufficiently stable across multiple cosmological epochs (Haas et al. 2003; Caplar et al. 2015, 2018). Large variations in candidate abundance or strong interval-dependent instabilities could indicate that the selection framework is dominated by redshift-dependent observational biases rather than intrinsic population properties. For this reason, the statistical pools were independently constructed within several redshift intervals, allowing the candidate distributions to be evaluated separately as functions of cosmological epoch. The resulting distributions indicate that statistically representative and cosmologically consistent candidates remain identifiable across all explored intervals. Although the number density of candidates varies between bins due to both observational completeness and intrinsic population evolution, the multidimensional pooling procedure consistently isolates quasars that reproduce the dominant observational properties of the corresponding parent populations (Pâris et al. 2018; Nardini et al. 2019; Bisogni et al. 2021). This persistence of statistically representative candidates across independent redshift intervals supports the robustness of the adopted selection framework and suggests that the procedure remains applicable over a relatively broad cosmological range. The complete catalog of the 255 photometric AGN candidates selected by this procedure is provided in Appendix~B.
\begin{figure}
\centering
\includegraphics[width=\linewidth]{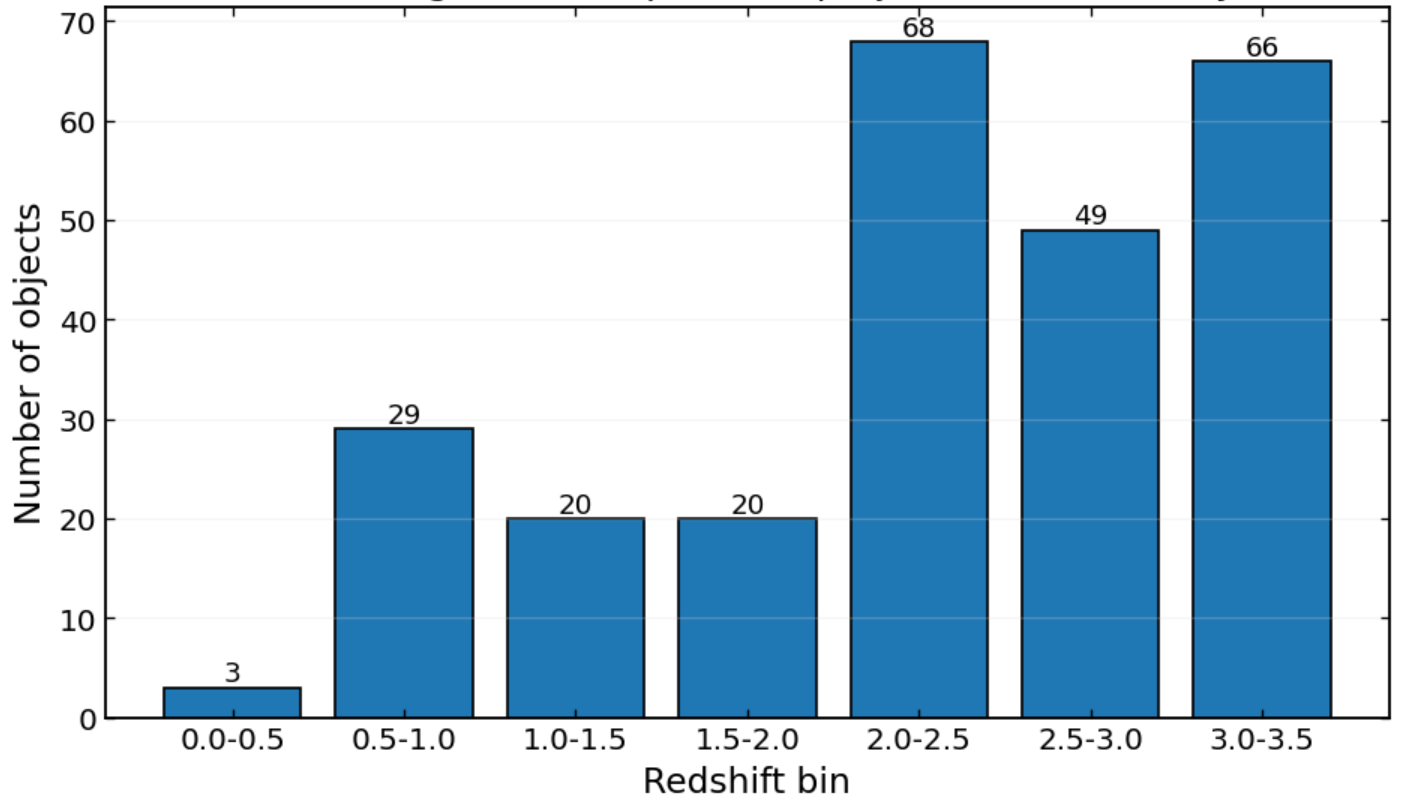}
\caption{
Redshift distribution of the photometric AGN golden sample (255) after applying the multidimensional pooling procedure and the projection-consistency criterion relative to the posterior-calibrated SN Ia-QSO cosmological reconstruction. The selected candidates remain present across all explored redshift intervals, indicating that statistically and cosmologically consistent AGN populations can be identified over a broad cosmological range.
}
\label{fig:redshift_distribution}
\end{figure}

Figure~\ref{fig:redshift_distribution} presents the redshift distribution of the final photometric AGN golden sample (255 objects) after applying both the multidimensional pooling procedure and the projection-consistency criterion introduced in Sect.~3.4. The statistical stability observed throughout the explored intervals further supports the interpretation that the multidimensional pooling methodology provides a reproducible mechanism for identifying cosmologically representative quasar populations within large photometric and spectroscopic surveys (Risaliti \& Lusso 2019; Lusso et al. 2020). The observed variation in the number of selected candidates across redshift intervals is not imposed by construction. Instead, it emerges naturally from the multidimensional statistical-distance criterion used during the selection process. In particular, the final candidate counts additionally depend on the adopted residual-consistency threshold, $|\Delta\mu| \leq 0.02$ mag, relative to the posterior-calibrated SN Ia--QSO cosmological reconstruction. Consequently, bins containing a larger number of candidates correspond to redshift intervals where a greater fraction of the parent population exhibits photometric properties compatible with the fiducial cosmological relation. The persistence of selected objects throughout the entire redshift range demonstrates that the procedure remains effective over a broad interval of cosmic time.

The final candidate distributions remain sufficiently populated throughout the explored cosmological intervals, suggesting that the statistical selection procedure remains stable over multiple epochs despite the intrinsic observational dispersion associated with quasar populations and the additional cosmological consistency requirement imposed by the residual-selection criterion (Nardini et al. 2019; Bisogni et al. 2021). The persistence of selected candidates throughout the explored redshift range indicates that the multidimensional pooling procedure remains effective across multiple cosmological epochs despite the intrinsic evolution of the parent quasar population. This behavior supports the robustness of the proposed framework for identifying statistically representative AGN candidates over a broad redshift interval.

\begin{figure*}[tbh]
\centering
\includegraphics[width=\linewidth]{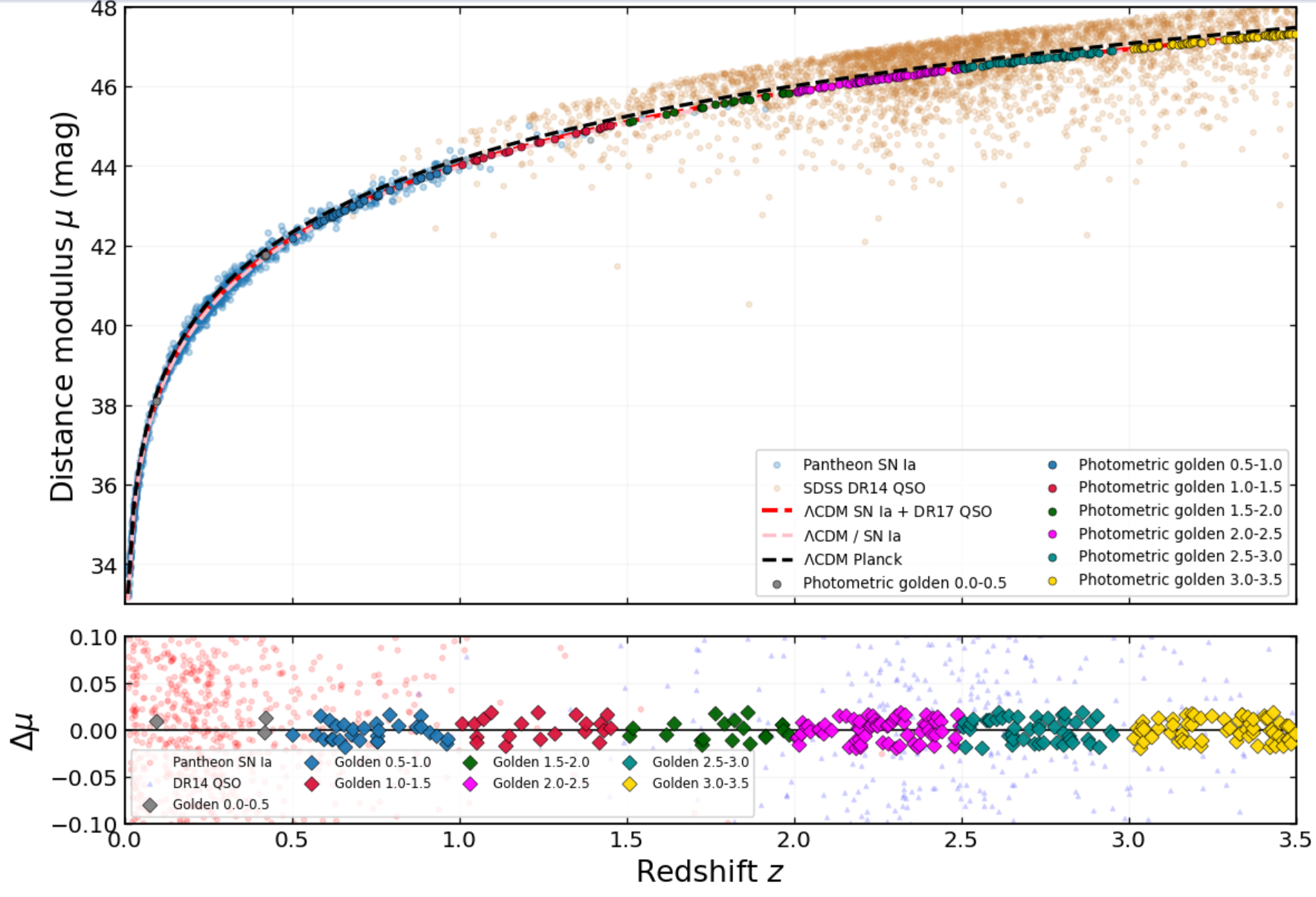}
\caption{Combined Type Ia supernova--quasar Hubble diagram including the SDSS DR17 photometric AGN golden sample. Colored symbols represent the statistically selected AGN candidates projected onto the fiducial cosmological relation inferred from the joint SN Ia--QSO calibration. The dashed curves show the posterior-calibrated $\Lambda$CDM model together with the comparison cosmologies. Lower panel: residual distance moduli relative to the fiducial cosmological reconstruction.}
\label{fig:hubble_golden}
\end{figure*}

\subsection{Reconstruction of the Hubble Diagram}

\begin{figure*}[p]
\centering

\includegraphics[width=0.6\linewidth]{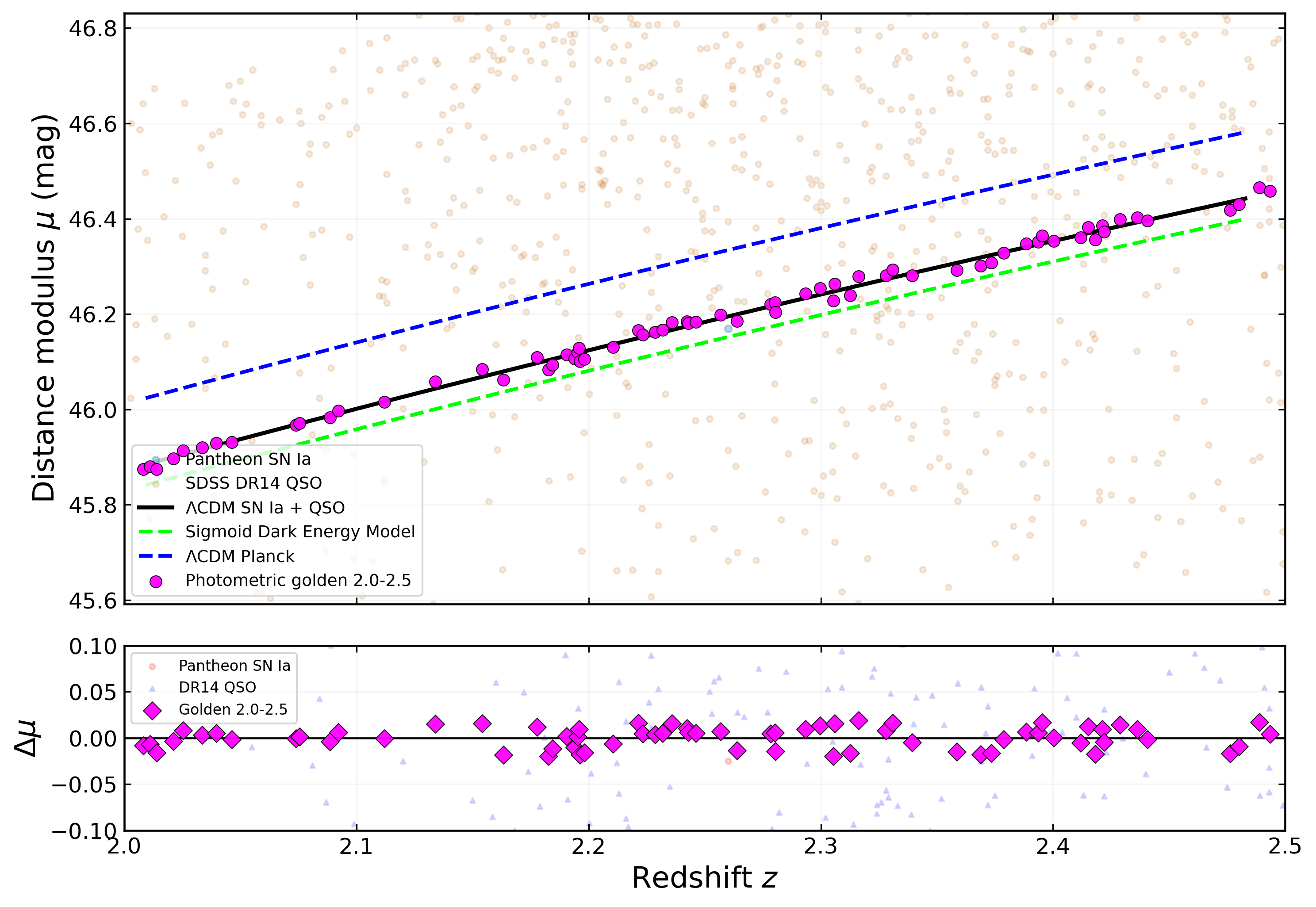}

\includegraphics[width=0.6\linewidth]{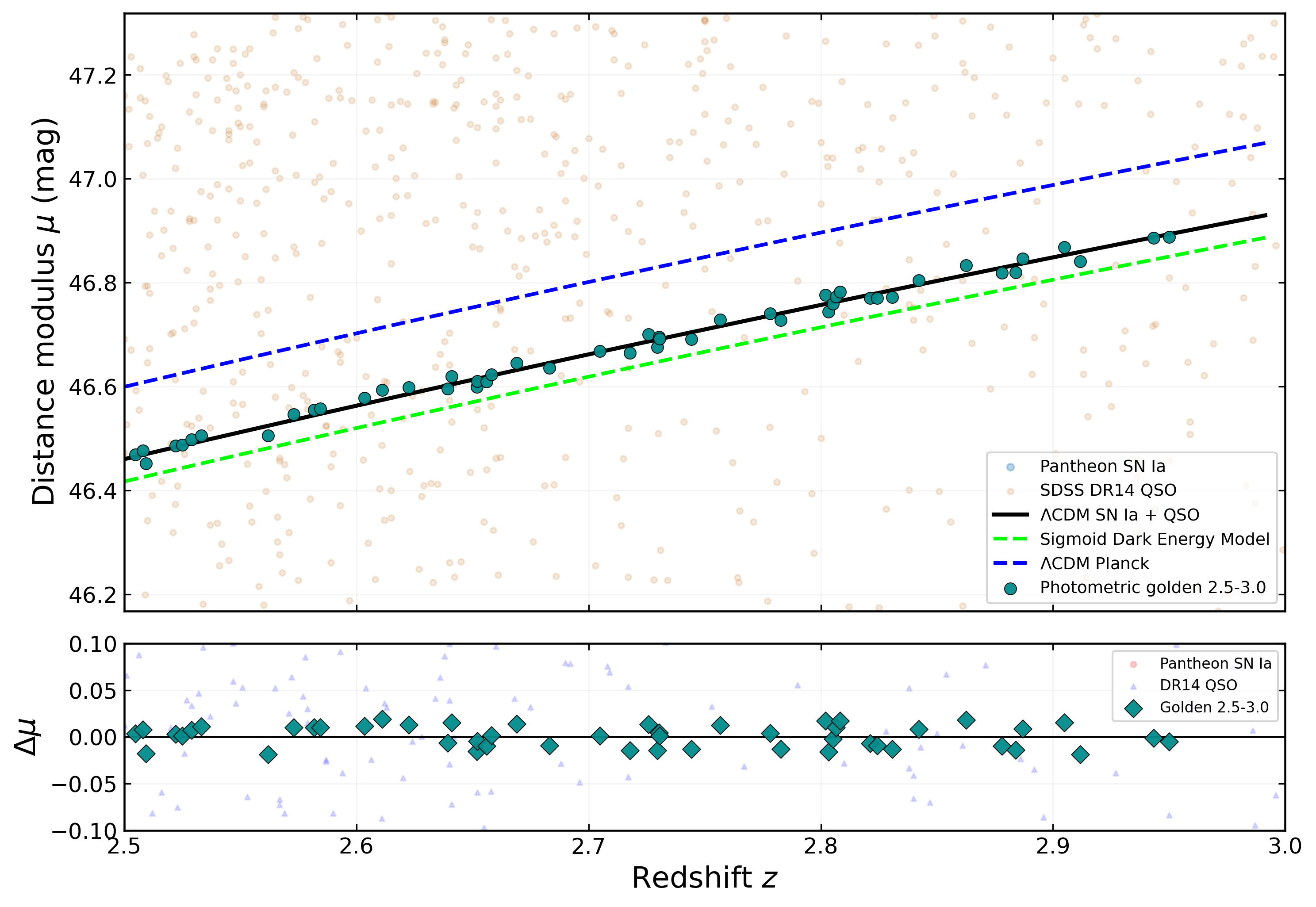}

\includegraphics[width=0.6\linewidth]{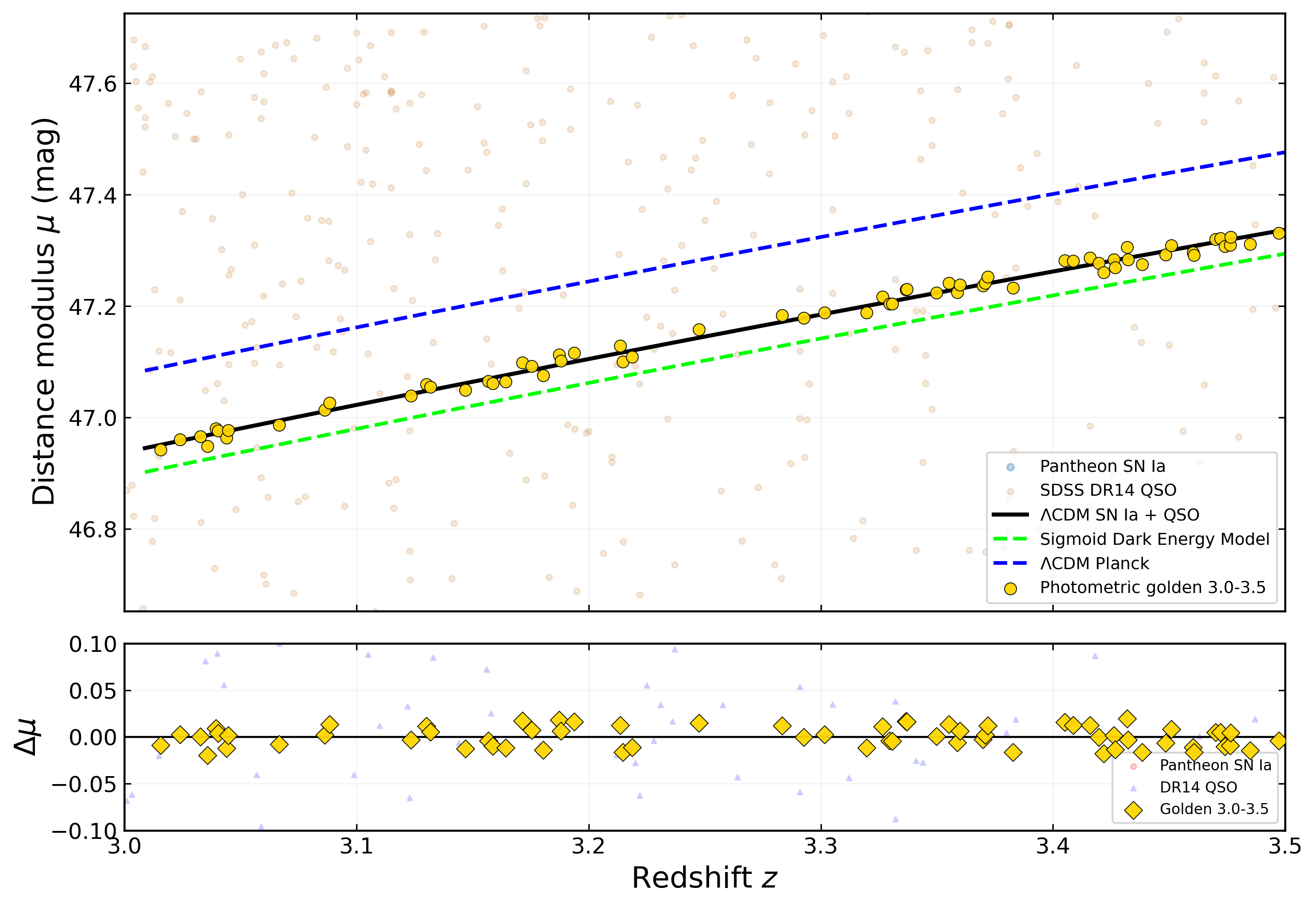}

\caption{
Zoomed plots of the combined Type Ia supernova-quasar Hubble diagram for the high-redshift intervals
(a) $2.0<z<2.5$, (b) $2.5<z<3.0$, and (c) $3.0<z<3.5$. The colored symbols represent the photometric AGN golden sample selected through the multidimensional pooling and cosmological-consistency criteria, while the parent SDSS DR14 quasar population is shown for comparison. The solid black curve corresponds to the posterior-calibrated SN Ia-QSO $\Lambda$CDM reconstruction. The lower panels display the residual distance moduli relative to the fiducial cosmological relation.
}
\label{fig:hubble_zoom_highz}
\end{figure*}

The statistically selected SDSS DR17 photometric candidates were projected onto the combined Type Ia supernova--quasar Hubble diagram in order to evaluate their cosmological consistency relative to the posterior-calibrated cosmological model. Within this framework, the Type Ia supernova sample defines the fiducial late-Universe cosmological relation, while the spectroscopic quasar population extends the observational coverage toward significantly higher redshifts (Risaliti \& Lusso 2019; Lusso et al. 2020; Zajaček et al. 2024). This combined approach allows the statistical behavior of the selected quasars to be directly compared against the cosmological reconstruction inferred from the posterior analysis. The selected quasar candidates were subsequently projected onto this common cosmological scale using the calibrated nuisance parameter $\Delta Q$ (Lusso \& Risaliti 2016; Risaliti \& Lusso 2019). A principal objective of this comparison is to determine whether the multidimensional statistical pooling procedure preferentially selects quasars that reproduce the dominant cosmological behavior traced by the supernova sample. In this context, the selected candidates are not interpreted as standard candles in the strict physical sense, but rather as statistically representative members of the parent quasar population whose observational properties remain consistent with the posterior-calibrated Hubble relation. The agreement persists despite the intrinsically larger dispersion associated with quasar populations relative to Type Ia supernovae (Nardini et al. 2019; Bisogni et al. 2021). Since quasars remain observable over a substantially broader redshift range, the statistical identification of cosmologically consistent candidates may provide a pathway toward constructing high-redshift cosmological tracers complementary to conventional supernova techniques (Risaliti \& Lusso 2019; Lusso et al. 2020; Zajaček et al. 2024). Figure~\ref{fig:hubble_golden} shows that the selected photometric AGN candidates remain compatible with the posterior-calibrated SN Ia--QSO cosmological reconstruction throughout the explored redshift range. Although the intrinsic dispersion of quasars exceeds that of Type Ia supernovae, the selected sample defines a comparatively narrow sequence around the fiducial Hubble relation, while the residuals remain centered on $\Delta\mu=0$ with no evident systematic trend as a function of redshift. Figure~\ref{fig:hubble_zoom_highz} presents zoomed reconstructions of the Hubble diagram for the intervals
$2.0<z<2.5$, $2.5<z<3.0$, and $3.0<z<3.5$. These
intervals contain the majority of the final photometric AGN
golden sample and therefore provide the most stringent test of
the selection procedure at epochs where Type Ia supernovae
become increasingly sparse.

The posterior median cosmology was subsequently used to reconstruct the combined Type Ia supernova--quasar Hubble diagram and evaluate the consistency of the statistically selected SDSS DR17 photometric candidates. The resulting reconstruction extends the observational Hubble relation toward significantly higher redshifts than those densely populated by the supernova sample alone. In this context, the selected quasars provide a preliminary high-redshift extension of the cosmological distance ladder \citep{Risaliti2019,Lusso2020}. The selected candidates reproduce the fiducial cosmological relation with relatively small residual dispersion across multiple redshift intervals. This behavior suggests that the multidimensional statistical pooling procedure preferentially isolates quasars whose observational properties remain consistent with the dominant cosmological behavior of the parent population. The enlarged high-redshift views presented in Figure~\ref{fig:hubble_zoom_highz} provide additional insight into the behavior of the selected candidates beyond the global reconstruction shown in Figure~\ref{fig:hubble_golden}, where the blue dashed lines corresponds to Planck distance modulus \citep{Planck2020} and the green dashed line corresponds to distance modulus calculated by using the Sigmoid DE Model \citep{TorresArzayus2024}.  Within these intervals, where the spectroscopic quasar population dominates the Hubble diagram and Type Ia supernovae become increasingly scarce, the photometric AGN golden sample emerges as a comparatively narrow sequence around the posterior-calibrated cosmological relation. This visual concentration indicates that the multidimensional pooling and projection-consistency criteria preferentially retain those objects whose observational properties remain statistically compatible with the fiducial reconstruction despite the intrinsically larger dispersion of the parent quasar population. The behavior observed further suggests that the proposed framework preserves the local structure of the Hubble diagram across independent redshift intervals rather than reproducing only its global trend. The persistence of this statistically coherent sequence at progressively higher redshifts supports the robustness of the selection methodology and reinforces the potential of the resulting photometric AGN golden sample as a foundation for future high-redshift cosmological analyses.

\subsection{Residual Dispersion and Statistical Stability}

Residual distance-modulus analyses were performed in order to investigate the statistical consistency of the selected quasar candidates relative to the posterior-calibrated cosmological reconstruction. Residual diagnostics constitute an important validation step because they allow systematic departures from the fiducial cosmological relation to be evaluated independently of the visual structure of the Hubble diagram itself. In particular, residual distributions provide direct information regarding the intrinsic dispersion of the selected quasar population and its stability across different redshift intervals. For each selected candidate, the residual distance modulus was computed relative to the posterior median cosmology inferred from the joint supernova--quasar analysis. The residual behavior indicates that the multidimensional statistical selection procedure, together with the projection-consistency requirement, preferentially isolates quasars whose projected distance moduli remain close to the fiducial cosmological relation. While the intrinsic scatter associated with quasars remains significantly larger than that of Type Ia supernovae, the selected candidates exhibit comparatively reduced residual dispersion relative to the full SDSS DR14 parent sample \citep{Nardini2019,Bisogni2021}. No evidence of strong systematic redshift-dependent deviations was identified within the explored cosmological intervals. This result suggests that the statistical pooling framework remains sufficiently stable across multiple epochs and does not artificially favor strongly biased subpopulations.

The residual root-mean-square (RMS) was computed as

\begin{equation}
{\rm RMS}(\Delta\mu)
=
\sqrt{
\frac{1}{N}
\sum_{i=1}^{N}
(\Delta\mu_i)^2
},
\end{equation}

where $\Delta\mu_i$ denotes the residual distance modulus of the $i$th candidate relative to the posterior-calibrated cosmological reconstruction.
\begin{table}[tbh]
\centering
\caption{Residual statistics of the photometric AGN golden sample in each redshift interval. The mean residual, residual dispersion, and root-mean-square (RMS) values remain approximately constant across the explored redshift range, indicating the stability of the multidimensional selection procedure.}
\label{tab:residual_stats}
\begin{tabular}{lcccc}
\hline
$z$ bin & $N$ & $\langle \Delta\mu \rangle$ (mag) & $\sigma_{\Delta\mu}$ (mag) & RMS (mag) \\
\hline
0.0-0.5 & 3 & 0.00671 & 0.00834 & 0.00956 \\
0.5-1.0 & 29 & -0.00183 & 0.00940 & 0.00941 \\
1.0-1.5 & 20 & 0.00233 & 0.01119 & 0.01116 \\
1.5-2.0 & 20 & -0.00070 & 0.01008 & 0.00985 \\
2.0-2.5 & 68 & 0.00041 & 0.01138 & 0.01130 \\
2.5-3.0 & 49 & 0.00076 & 0.01170 & 0.01160 \\
3.0-3.5 & 66 & 0.00029 & 0.01101 & 0.01093 \\
\hline
\end{tabular}
\end{table}

The relatively small residual dispersion observed for the final photometric AGN golden sample further supports the applicability of the combined pooling and projection-consistency framework for identifying cosmologically representative quasar populations. The statistics reported in Table~\ref{tab:residual_stats} indicate that the residual distributions remain stable across the explored redshift range. The residual means are consistent with zero, while both the dispersion and RMS remain approximately constant, suggesting that the multidimensional pooling procedure does not introduce significant redshift-dependent biases. Figure~5 provides a visual confirmation of this behavior, showing no systematic drift of the residuals with increasing redshift.

\begin{figure}[tbh]
\centering
\includegraphics[width=0.7\linewidth]{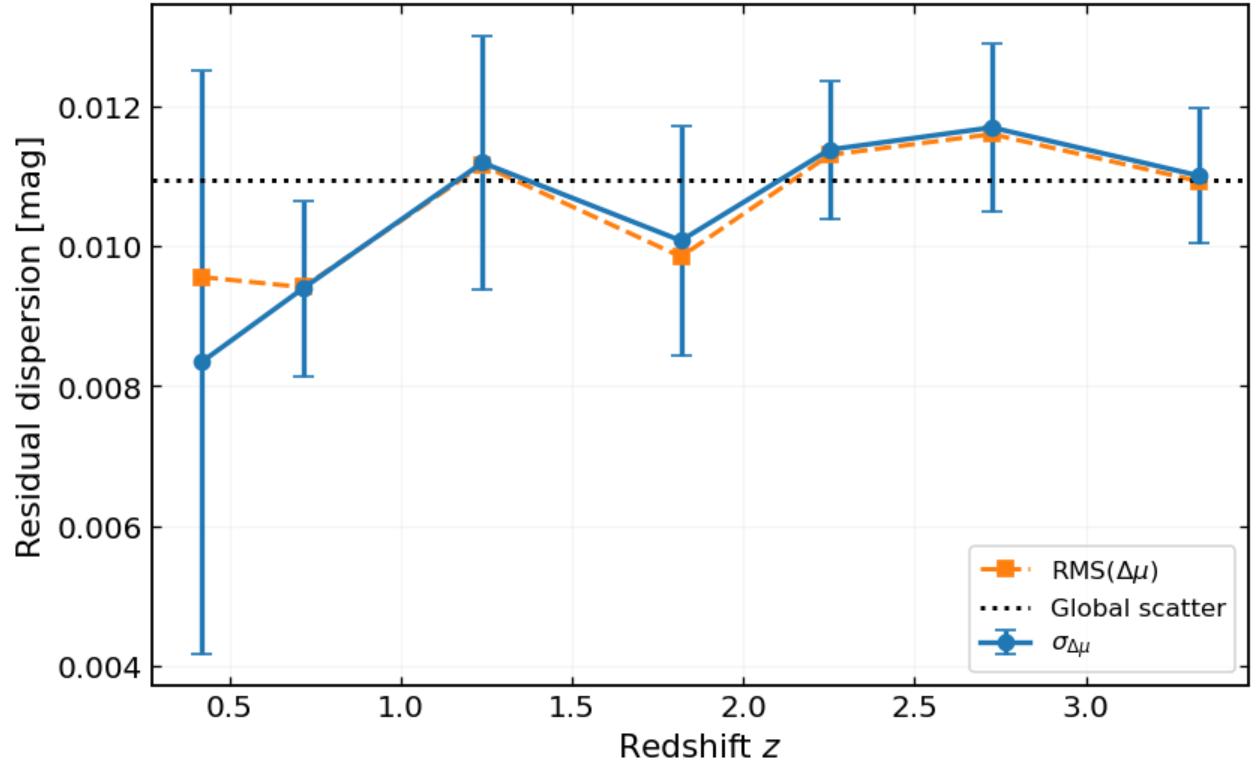}
\caption{
Residual-dispersion statistics of the photometric AGN golden sample as a function of redshift. The RMS residuals and the residual standard deviations remain approximately constant across the explored redshift range, indicating statistical stability of the selection procedure and consistency with the posterior-calibrated cosmological reconstruction.}
\label{fig:residuals}
\end{figure}

Figure~\ref{fig:residuals} further illustrates the statistical consistency of the selected candidates. The residuals remain symmetrically distributed around $\Delta\mu = 0$ with no visible systematic drift as a function of redshift $z$. The absence of large-scale trends in the residual distribution is particularly important. If the multidimensional distance metric were introducing a redshift-dependent selection bias, one would expect the residuals to exhibit coherent positive or negative excursions in specific redshift intervals. Instead, the selected candidates remain approximately centered on the fiducial relation from the nearby Universe to $z\approx3.5$. The resulting behavior suggests that the pooling procedure is not merely selecting objects with similar redshifts or magnitudes, but rather identifying quasars whose combined observational properties reproduce the dominant cosmological behavior encoded in the calibrated supernova--quasar Hubble relation.

The statistical stability demonstrated by the residual analysis has implications beyond the internal validation of the selection procedure. The absence of systematic trends with redshift indicates that the multidimensional pooling framework preserves the global structure of the calibrated SN Ia--QSO Hubble relation while substantially reducing the observational dispersion relative to the parent quasar population. This behavior suggests that the selected candidates are not simply characterized by similar photometric properties, but instead represent objects whose combined observational characteristics remain mutually consistent over a broad cosmological baseline.

From a methodological perspective, these results support the use of multidimensional statistical criteria as an effective strategy for identifying promising AGN candidates for cosmological applications. Rather than attempting to standardize the entire quasar population, the present framework isolates a restricted subset whose statistical behavior is significantly more homogeneous than that of the parent sample. Although this statistical coherence does not by itself imply physical homogeneity, it provides a natural starting point for subsequent physical characterization through spectral energy distribution modeling and multiwavelength analyses. In this sense, the residual analysis presented here constitutes not only a validation of the selection procedure but also the statistical foundation upon which future AGN standardization efforts may be developed.

\subsection{Prospects for AGN Standardization and Cosmological Applications}

The residual analyses presented above motivate a broader interpretation of the proposed selection framework. Beyond reproducing the calibrated SN Ia--QSO Hubble relation, the statistical behavior of the photometric AGN golden sample raises the possibility that the selected objects represent a physically more homogeneous subset of the parent quasar population. By simultaneously combining redshift, broadband photometry, color information, and compatibility with the posterior-calibrated SN Ia--QSO cosmological reconstruction, the methodology isolates candidates that exhibit reduced dispersion with respect to the fiducial Hubble relation. Although this statistical coherence does not by itself demonstrate physical homogeneity, it provides a strong indication that the selected objects may occupy a restricted region of the multidimensional parameter space governing AGN emission and accretion processes \citep{Risaliti2015,Risaliti2019,Lusso2020,Bisogni2021}. Unlike traditional AGN selection strategies based primarily on luminosity, signal-to-noise ratio, or spectroscopic quality, the present framework seeks the most statistically representative members of the parent population. The underlying assumption is that reducing the diversity of the observed population may naturally reduce the intrinsic scatter of the UV--X-ray luminosity relation without imposing explicit physical constraints during the selection stage. Consequently, the proposed methodology should be regarded as a statistically guided identification procedure whose physical interpretation must be established through independent observational analyses rather than assumed a priori.

If the selected quasars indeed constitute a physically homogeneous population, one may expect them to exhibit comparable black-hole masses, Eddington ratios, accretion efficiencies, disk temperatures, and disk--corona coupling properties. Likewise, reduced dispersion in bolometric corrections, dust attenuation, orientation effects, and host-galaxy contamination could naturally contribute to the stability of the UV--X-ray luminosity relation and therefore to the observed reduction in cosmological scatter \citep{Nardini2019,Haas2003,Sacchi2022}. These hypotheses remain speculative at the present stage and require direct observational verification. Such verification will require detailed physical characterization of the selected AGNs through spectral energy distribution fitting and multiwavelength observations spanning the ultraviolet, optical, infrared, and X-ray regimes. These analyses will provide estimates of bolometric luminosities, accretion rates, black-hole masses, host-galaxy contributions, dust content, and other physical parameters that can be directly compared with those of the parent quasar population. Establishing whether the statistically selected sample also forms a physically homogeneous subclass represents the natural continuation of the present work and an essential step toward the eventual standardization of AGNs as precision cosmological probes \citep{Bisogni2021,Sacchi2022,Zajacek2024}.

From a cosmological perspective, the methodology presented here is intentionally independent of detailed physical modeling during the initial selection stage. Instead, it provides an objective, reproducible, and computationally efficient framework for identifying promising AGN candidates within very large photometric surveys. As forthcoming facilities are expected to increase the number of known AGNs by several orders of magnitude, reproducible statistical selection techniques will become increasingly valuable for defining high-quality samples suitable for subsequent spectroscopic characterization and physical calibration. In this sense, the photometric AGN golden sample presented in this work should be viewed as an intermediate step toward constructing physically validated AGN samples capable of extending precision cosmology well beyond the redshift range currently accessible to Type Ia supernovae \citep{Caplar2015,Caplar2018,Zajacek2024}.

\section{Comparison with other values of $H_{0}$}

The Hubble constant inferred from the joint SN Ia--QSO analysis
can be compared with a variety of independent determinations
reported in the recent literature. Such comparisons provide an
additional consistency test for the cosmological reconstruction
obtained in this work and allow the inferred value to be placed
within the broader context of the current Hubble-tension debate.

This intermediate position is particularly interesting because
the SN Ia--QSO reconstruction combines a low-redshift
supernova anchor with a high-redshift quasar extension,
thereby probing a cosmological regime that differs from those
traditionally employed in local distance-ladder analyses.
While the present work is not intended to resolve the Hubble
tension, the resulting value demonstrates that quasar-based
cosmological reconstructions can provide competitive and
independent constraints on the expansion rate of the Universe.

\begin{figure*}
\centering
\includegraphics[width=\linewidth]{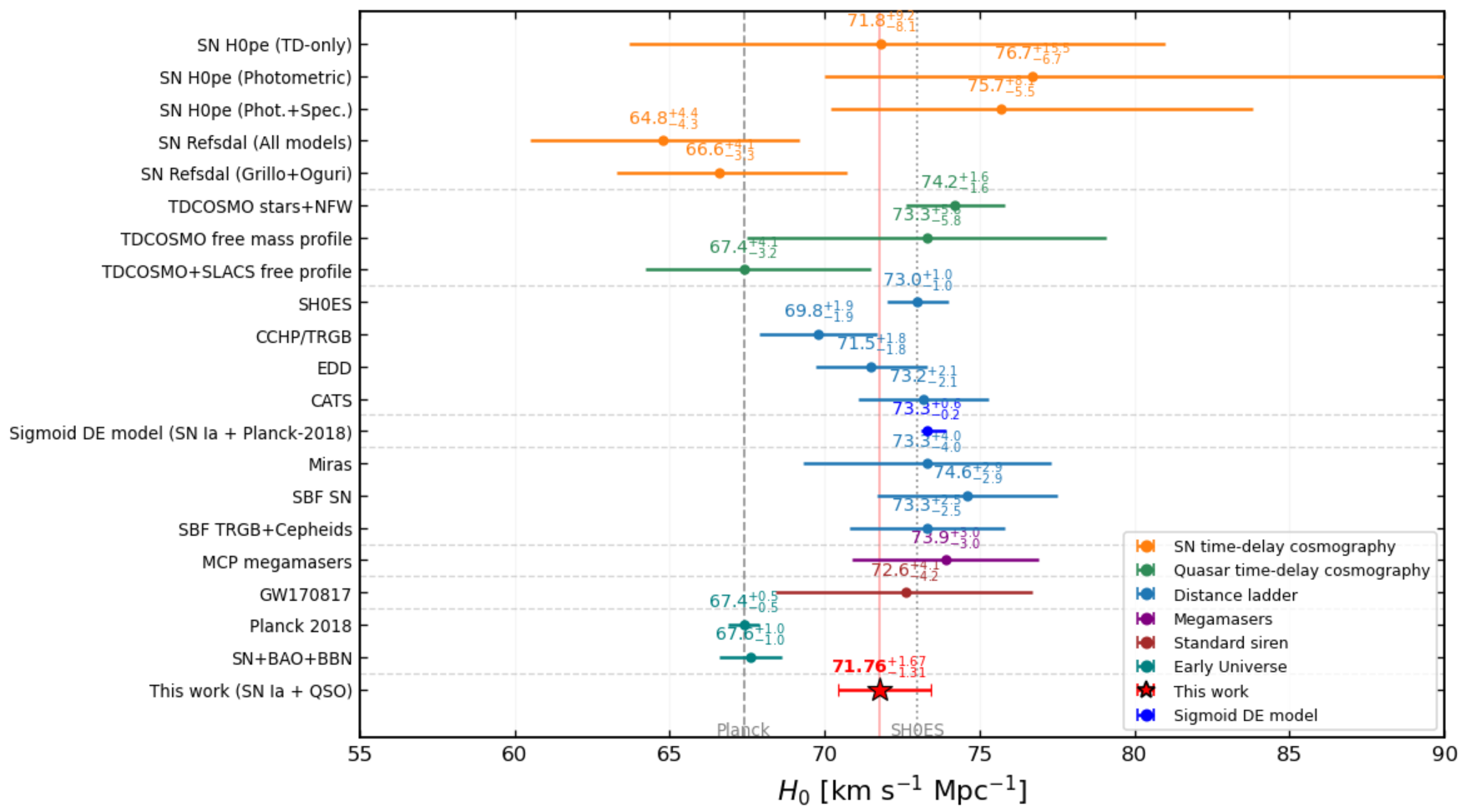}
\caption{
Comparison between the Hubble constant obtained in this work and representative determinations reported in the recent literature. The compiled measurements include the three determinations from SN H0pe (time-delay only, photometric, and combined photometric plus spectroscopic analyses; \citealt{Pascale2025}), the two determinations from SN Refsdal (all lens models and the preferred Grillo-g plus Oguri-a* models; \citealt{Kelly2023}), TDCOSMO (stars+NFW, free mass-profile, and TDCOSMO+SLACS free mass-profile analyses; \citealt{Millon2020,Birrer2020,Shajib2023}), SH0ES (\citealt{Riess2022}), CCHP/TRGB (\citealt{Freedman2019}), EDD (\citealt{Anand2022}), CATS (\citealt{Scolnic2023}), Miras (\citealt{Huang2020}), SBF calibrated through Type Ia supernovae (\citealt{Garnavich2023}), SBF calibrated with TRGB and Cepheids (\citealt{Blakeslee2021}), MCP megamasers (\citealt{Pesce2020}), GW170817 standard sirens (\citealt{Wang2023}), Planck 2018 (\citealt{Planck2020}), Sigmoid DE model (\citealt{TorresArzayus2024})  and the joint SN+BAO+BBN determination (\citealt{Schoeneberg2022}). Horizontal bars indicate the reported $1\sigma$ uncertainties. The red star denotes the joint SN Ia--QSO determination obtained in the present work. The literature compilation follows the presentation of \citet{Pascale2025}, while the corresponding original references are indicated for each independent determination.
}
\label{fig:H0_comparison}
\end{figure*}

From a cosmological perspective, this result is noteworthy
because it is obtained through a methodology that differs
substantially from both traditional distance-ladder approaches
and early-Universe probes. The present reconstruction relies
on the extension of the Hubble diagram to redshifts
significantly higher than those directly accessible through
Type Ia supernova observations, exploiting the empirical
$L_X-L_{UV}$ relation of quasars as an independent distance
indicator. Consequently, the agreement with several late-time
measurements suggests that quasar-based cosmology can
provide complementary information on the expansion history
of the Universe over an extended redshift range. Although the
precision achieved here does not allow a definitive assessment
of the Hubble tension, the consistency of the inferred value
supports the continued development of quasar samples as
high-redshift cosmological probes (Lusso \& Risaliti 2016;
Risaliti \& Lusso 2019; Lusso et al. 2020). Future improvements in the selection of cosmologically
reliable quasars, together with forthcoming high-quality
spectroscopic and multiwavelength observations, are expected
to reduce the intrinsic scatter of the $L_X-L_{UV}$ relation and
therefore improve the precision of quasar-based constraints
on $H_0$. In this context, the photometric AGN golden sample
identified in the present work represents a promising step
toward the construction of increasingly robust high-redshift
Hubble diagrams.

The comparison presented in Figure~\ref{fig:H0_comparison}
shows that the value inferred from the joint SN Ia--QSO
reconstruction occupies an intermediate position between the
early-Universe determination obtained from measurements of
the cosmic microwave background and several late-Universe
distance-ladder estimates. In particular, the posterior value
derived in this work,
$H_0 = 71.76^{+1.67}_{-1.31}\,\mathrm{km\,s^{-1}\,Mpc^{-1}}$,
is higher than the value reported by the \textit{Planck}
Collaboration assuming the standard $\Lambda$CDM model
(\(H_0 = 67.4 \pm 0.5\) km s$^{-1}$ Mpc$^{-1}$;
Planck Collaboration VI 2020), while remaining compatible, within the quoted uncertainties, with several independent late-time determinations based on Cepheids and Type Ia
supernovae (Riess et al. 2022), strong-lensing time-delay cosmography (Millon et al. 2020; Wong et al. 2020),
tip-of-the-red-giant-branch calibrations (Freedman et al. 2021), megamaser observations (Pesce et al. 2020), and
gravitational-wave standard sirens (Abbott et al. 2017). Although the uncertainty associated with the present determination remains larger than that achieved by the most mature cosmological techniques, the agreement with multiple independent methods demonstrates that the proposed photometric AGN selection does not introduce significant systematic shifts into the inferred cosmological expansion rate. This consistency provides an important validation of the multidimensional selection framework developed throughout this work. More importantly, the significance of the present result extends beyond the numerical estimate of $H_0$. The methodology introduced here establishes a reproducible procedure for identifying statistically consistent AGN populations capable of extending cosmological analyses to redshifts substantially beyond the range directly accessible to Type Ia supernovae. As increasingly large photometric and spectroscopic surveys become available, improvements in sample size, photometric quality, and physical characterization are expected to reduce both statistical and systematic uncertainties. Consequently, the framework presented here should be regarded as a first step toward the construction of high-redshift AGN samples suitable for future precision-cosmology applications, where independent determinations of the cosmic expansion history will become increasingly valuable. The agreement observed therefore represents not only a consistency check of the inferred Hubble constant, but also an independent validation of the statistical methodology used to construct the photometric AGN golden sample.

\section{Conclusions}

We have presented a joint Type Ia supernova--quasar cosmological analysis aimed at developing a reproducible framework for the identification of statistically and cosmologically consistent photometric AGN candidates. By combining the Pantheon+SH0ES supernova compilation with SDSS DR14 quasars, we constructed a common Hubble diagram and inferred the posterior distributions of the cosmological and nuisance parameters within a flat $\Lambda$CDM framework. The resulting posterior medians, $H_0 \simeq 71.9$ km s$^{-1}$ Mpc$^{-1}$ and $\Omega_m \simeq 0.295$, were subsequently adopted as the fiducial cosmological reference for the photometric selection procedure.

Using SDSS DR17 photometric data, we constructed a parent pool of AGN candidates spanning the redshift interval $0<z<3.5$ and defined a multidimensional statistical-distance metric based on redshift, optical magnitudes, and SDSS colors. This metric was used to identify candidates occupying the central region of the photometric parameter space. An additional cosmological-consistency criterion based on the residual distance modulus relative to the posterior-calibrated Hubble relation was then applied in order to isolate those objects whose projected cosmological behavior remains compatible with the fiducial reconstruction. The resulting photometric AGN golden sample exhibits a high degree of statistical consistency with the reference cosmological relation. The residual bias remains compatible with zero, the residual dispersion shows no evidence of strong redshift-dependent evolution, and the selected candidates reproduce the posterior-calibrated Hubble diagram throughout the explored redshift interval. Furthermore, the stability of the residual statistics across multiple cosmological epochs supports the robustness of the combined multidimensional pooling and projection-consistency framework. For reproducibility, the mathematical derivation of the statistical framework is summarized in Appendix~A, while the complete photometric AGN golden sample is reported in Appendix~B.

The present work does not establish the physical standardization of the selected AGN population. Instead, it introduces a reproducible statistical methodology for identifying photometric AGN candidates whose observational properties remain simultaneously consistent with both the dominant characteristics of the parent population and a calibrated cosmological reconstruction. In this sense, the primary result of this study is not the construction of a final standardized AGN sample, but the development of a systematic procedure for identifying promising candidates for future physical calibration.

Because active galactic nuclei can be observed at redshifts substantially higher than those routinely accessible to Type Ia supernovae, the methodology developed here may provide a pathway toward extending cosmological distance-indicator studies into previously unexplored regions of cosmic history. Future investigations involving spectral energy distribution fitting, multiwavelength observations, and detailed characterization of the selected candidates will be necessary to determine whether the statistical homogeneity identified in this work corresponds to an underlying physical population capable of supporting AGN standardization efforts. The framework presented here therefore constitutes a first step toward the construction of physically motivated high-redshift AGN samples for precision cosmology.







   
  



\appendix
\section{Mathematical derivation of the cosmological and statistical framework}

This appendix summarizes the main mathematical steps underlying the cosmological reconstruction and the statistical selection procedure described in Sects.~2 and 3. The goal is to provide the derivations needed to connect the luminosity-distance formalism, the quasar UV--X-ray relation, the joint likelihood, and the multidimensional distance criterion used to define the photometric AGN golden sample.

\subsection{Luminosity distance and distance modulus}

For a spatially flat FLRW metric, the radial propagation of photons satisfies

\begin{equation}
ds^2=0,
\qquad
d\Omega=0,
\end{equation}

so that

\begin{equation}
dr=\frac{c\,dt}{a(t)}.
\end{equation}

Using $a=(1+z)^{-1}$ and $H=\dot a/a$, one obtains

\begin{equation}
dt=-\frac{dz}{(1+z)H(z)}.
\end{equation}

Therefore,

\begin{equation}
dr=-\frac{c\,dz}{H(z)}.
\end{equation}

Integrating from emission to observation gives the comoving radial distance

\begin{equation}
r(z)=c\int_0^z\frac{dz'}{H(z')}.
\end{equation}

For a flat universe, the luminosity distance is

\begin{equation}
d_L(z)=(1+z)r(z),
\end{equation}

and therefore

\begin{equation}
d_L(z)
=
c(1+z)
\int_0^z
\frac{dz'}{H(z')}.
\end{equation}

For flat $\Lambda$CDM,

\begin{equation}
H(z)
=
H_0
\left[
\Omega_m(1+z)^3+1-\Omega_m
\right]^{1/2},
\end{equation}

which yields

\begin{equation}
d_L(z)
=
\frac{c(1+z)}{H_0}
\int_0^z
\frac{dz'}
{
\left[
\Omega_m(1+z')^3+1-\Omega_m
\right]^{1/2}
}.
\end{equation}

The distance modulus follows from the inverse-square flux relation,

\begin{equation}
F=\frac{L}{4\pi d_L^2}.
\end{equation}

Since

\begin{equation}
m-M=-2.5\log_{10}
\left(
\frac{F}{F_{10{\rm pc}}}
\right),
\end{equation}

and

\begin{equation}
\frac{F}{F_{10{\rm pc}}}
=
\frac{(10\,{\rm pc})^2}{d_L^2},
\end{equation}

one obtains

\begin{equation}
\mu
\equiv
m-M
=
5\log_{10}
\left(
\frac{d_L}{10\,{\rm pc}}
\right).
\end{equation}

Expressing $d_L$ in Mpc gives

\begin{equation}
\mu(z)
=
5\log_{10}
\left[
\frac{d_L(z)}{\rm Mpc}
\right]
+
25.
\end{equation}

\subsection{Supernova likelihood}

For Type Ia supernovae, the observed distance modulus is

\begin{equation}
\mu_{{\rm SN},i}
=
m_{B,i}-M_B,
\end{equation}

and the model prediction is

\begin{equation}
\mu_{\rm th}(z_i)
=
5\log_{10}
\left[
\frac{d_L(z_i;H_0,\Omega_m)}{\rm Mpc}
\right]
+
25.
\end{equation}

The residual for each supernova is

\begin{equation}
R_{{\rm SN},i}
=
\mu_{{\rm SN},i}
-
\mu_{\rm th}(z_i).
\end{equation}

Assuming independent Gaussian uncertainties, the likelihood is

\begin{equation}
\mathcal{L}_{\rm SN}
\propto
\exp
\left[
-\frac{1}{2}
\chi^2_{\rm SN}
\right],
\end{equation}

with

\begin{equation}
\chi^2_{\rm SN}
=
\sum_i
\frac{
R_{{\rm SN},i}^2
}
{\sigma_{{\rm SN},i}^2}.
\end{equation}

\subsection{Quasar UV--X-ray relation and distance scale}

The quasar luminosity relation is

\begin{equation}
\log L_X
=
\gamma\log L_{\rm UV}+\beta .
\end{equation}

Using

\begin{equation}
L_X=4\pi d_L^2F_X,
\qquad
L_{\rm UV}=4\pi d_L^2F_{\rm UV},
\end{equation}

one obtains

\begin{equation}
\log F_X+\log(4\pi)+2\log d_L
=
\gamma
\left[
\log F_{\rm UV}+\log(4\pi)+2\log d_L
\right]
+
\beta .
\end{equation}

Collecting terms in $\log d_L$ gives

\begin{equation}
2(1-\gamma)\log d_L
=
\gamma\log F_{\rm UV}
-
\log F_X
+
(\gamma-1)\log(4\pi)
+
\beta .
\end{equation}

Thus,

\begin{equation}
\log d_L
=
\frac{
\gamma\log F_{\rm UV}
-
\log F_X
+
(\gamma-1)\log(4\pi)
+
\beta
}
{2(1-\gamma)}.
\end{equation}

The corresponding quasar distance modulus can be written as

\begin{equation}
\mu_{\rm QSO}
=
\frac{5}{2(1-\gamma)}
\left[
\gamma\log F_{\rm UV}
-
\log F_X
\right]
+
\mathcal{C},
\end{equation}

where $\mathcal{C}$ absorbs the intercept $\beta$, the factor $4\pi$, unit conversions, and the absolute normalization of the UV--X-ray relation.

Since this additive normalization cannot be determined from fluxes alone, it is represented in the joint reconstruction by the quasar offset parameter $\Delta Q$. The quasar residual is therefore modeled as

\begin{equation}
R_{{\rm QSO},j}
=
\mu_{{\rm QSO},j}
-
\mu_{\rm th}(z_j)
-
\Delta Q .
\end{equation}

The parameter $\Delta Q$ is not an intrinsic quasar luminosity. Instead, it is an effective population-level calibration term that places the quasar distance indicator on the same distance-modulus scale as the Type Ia supernova sample.

\subsection{Quasar likelihood and joint posterior}

Assuming Gaussian quasar residuals, the quasar likelihood is

\begin{equation}
\mathcal{L}_{\rm QSO}
\propto
\exp
\left[
-\frac{1}{2}
\chi^2_{\rm QSO}
\right],
\end{equation}

with

\begin{equation}
\chi^2_{\rm QSO}
=
\sum_j
\frac{
R_{{\rm QSO},j}^2
}
{\sigma_{{\rm QSO},j}^2}.
\end{equation}

Assuming that the supernova and quasar datasets are statistically independent,

\begin{equation}
\mathcal{L}_{\rm tot}
=
\mathcal{L}_{\rm SN}
\mathcal{L}_{\rm QSO}.
\end{equation}

Therefore,

\begin{equation}
\mathcal{L}_{\rm tot}
\propto
\exp
\left[
-\frac{1}{2}
\left(
\chi^2_{\rm SN}
+
\chi^2_{\rm QSO}
\right)
\right],
\end{equation}

or

\begin{equation}
\chi^2_{\rm tot}
=
\chi^2_{\rm SN}
+
\chi^2_{\rm QSO}.
\end{equation}

The posterior distribution is then

\begin{equation}
P(\Theta|D)
\propto
\mathcal{L}_{\rm tot}(D|\Theta)P(\Theta),
\end{equation}

where

\begin{equation}
\Theta=(H_0,\Omega_m,M_B,\Delta Q).
\end{equation}

The posterior median values define the fiducial cosmological relation used to project and evaluate the photometric AGN candidates.

\subsection{Projection residuals}

For the posterior median cosmology, the projected distance modulus is

\begin{equation}
\mu_{\rm proj}^{\rm posterior}(z)
=
5\log_{10}
\left[
\frac{
d_L(z;H_0^{\rm med},\Omega_m^{\rm med})
}
{\rm Mpc}
\right]
+
25.
\end{equation}

The residual associated with each photometric AGN candidate is

\begin{equation}
\Delta\mu_{\rm posterior}
=
\mu_{\rm candidate}
-
\mu_{\rm proj}^{\rm posterior}.
\end{equation}

The final projection-consistency condition is

\begin{equation}
|\Delta\mu_{\rm posterior}|
\leq
0.02~{\rm mag}.
\end{equation}

The magnitude residual can be related to a fractional distance deviation. Since

\begin{equation}
\delta\mu
=
5\log_{10}
\left(
1+\frac{\delta d_L}{d_L}
\right),
\end{equation}

for $\delta d_L/d_L\ll1$,

\begin{equation}
\frac{\delta d_L}{d_L}
\simeq
\frac{\ln 10}{5}\delta\mu.
\end{equation}

For $\delta\mu=0.02$ mag,

\begin{equation}
\frac{\delta d_L}{d_L}
\simeq
9.2\times10^{-3},
\end{equation}

so the adopted residual window corresponds to a fractional luminosity-distance deviation of order one percent.

\subsection{Multidimensional statistical distance}

The photometric candidate selection is performed in the observable space

\begin{equation}
\mathbf{x}
=
(z,u,g,r,i,z_{\rm SDSS},u-g,g-r,r-i,i-z_{\rm SDSS}).
\end{equation}

For each redshift pool, the central vector is

\begin{equation}
\tilde{\mathbf{x}}
=
(\tilde{x}_1,\tilde{x}_2,\ldots,\tilde{x}_n),
\end{equation}

where $\tilde{x}_i$ is the median of the $i$th observable. The general Mahalanobis distance is

\begin{equation}
D_M^2
=
(\mathbf{x}-\tilde{\mathbf{x}})^T
\mathbf{C}^{-1}
(\mathbf{x}-\tilde{\mathbf{x}}).
\end{equation}

Under the diagonal covariance approximation,

\begin{equation}
\mathbf{C}
\simeq
{\rm diag}
(\sigma_1^2,\sigma_2^2,\ldots,\sigma_n^2),
\end{equation}

and therefore

\begin{equation}
\mathbf{C}^{-1}
\simeq
{\rm diag}
\left(
\sigma_1^{-2},
\sigma_2^{-2},
\ldots,
\sigma_n^{-2}
\right).
\end{equation}

Substitution gives

\begin{equation}
D_{\rm phot}^{2}
=
\sum_{i=1}^{n}
\left(
\frac{x_i-\tilde{x}_i}{\sigma_i}
\right)^2.
\end{equation}

Thus, each observable contributes in units of its own dispersion. Sources with small $D_{\rm phot}^{2}$ are located near the multidimensional center of the parent population, while large values correspond to increasingly atypical objects.

\subsection{Residual dispersion statistics}

For a set of $N$ selected candidates, the mean residual is

\begin{equation}
\langle\Delta\mu\rangle
=
\frac{1}{N}
\sum_{i=1}^{N}
\Delta\mu_i.
\end{equation}

The residual standard deviation is

\begin{equation}
\sigma_{\Delta\mu}
=
\left[
\frac{1}{N-1}
\sum_{i=1}^{N}
(\Delta\mu_i-\langle\Delta\mu\rangle)^2
\right]^{1/2},
\end{equation}

and the root-mean-square residual is

\begin{equation}
{\rm RMS}(\Delta\mu)
=
\left[
\frac{1}{N}
\sum_{i=1}^{N}
(\Delta\mu_i)^2
\right]^{1/2}.
\end{equation}

Expanding $\Delta\mu_i=(\Delta\mu_i-\langle\Delta\mu\rangle)+\langle\Delta\mu\rangle$, one obtains

\begin{equation}
{\rm RMS}^2(\Delta\mu)
\simeq
\sigma_{\Delta\mu}^2
+
\langle\Delta\mu\rangle^2,
\end{equation}

where the approximation becomes exact up to the usual $N/(N-1)$ variance normalization factor. Therefore, when the mean residual is close to zero,

\begin{equation}
{\rm RMS}(\Delta\mu)
\simeq
\sigma_{\Delta\mu}.
\end{equation}

This relation explains why stable RMS and standard-deviation values across redshift intervals indicate that the selected candidates remain statistically consistent with the posterior-calibrated Hubble relation.

\subsection{Compact mathematical workflow}

For each redshift bin $B_k$, the parent pool is

\begin{equation}
\mathcal{P}_k
=
\{
\mathbf{x}_j:\ z_j\in B_k
\}.
\end{equation}

The median and dispersion of each observable are computed as

\begin{equation}
\tilde{x}_{i,k}
=
{\rm median}(x_{i,j})_{j\in\mathcal{P}_k},
\qquad
\sigma_{i,k}
=
{\rm std}(x_{i,j})_{j\in\mathcal{P}_k}.
\end{equation}

For each candidate,

\begin{equation}
D_{{\rm phot},j,k}^{2}
=
\sum_i
\left(
\frac{x_{i,j}-\tilde{x}_{i,k}}
{\sigma_{i,k}}
\right)^2.
\end{equation}

The final golden sample is obtained by retaining statistically central and cosmologically consistent candidates,

\begin{equation}
\mathcal{G}
=
\left\{
j:
D_{{\rm phot},j,k}^{2}\ {\rm small},
\quad
|\Delta\mu_{{\rm posterior},j}|
\leq
0.02~{\rm mag}
\right\}.
\end{equation}

This compact workflow summarizes the mathematical construction of the photometric AGN golden sample used in the main analysis.

\section{Photometric AGN Golden Sample}

This appendix presents the complete catalog of the photometric AGN golden sample identified through the multidimensional statistical framework developed in this work. The catalog contains the 255 quasars selected after applying the multidimensional pooling analysis, the posterior projection onto the joint SN Ia--QSO cosmological reconstruction, and the projection-consistency criterion described in Sects.~2--4.

To preserve a compact and readable printed layout, the catalog is reported using the SDSS object identifier as the main source identifier. The sexagesimal coordinates are omitted from the printed table, while the decimal equatorial coordinates, redshift, redshift interval, statistical distance, projected distance modulus, posterior-calibrated distance modulus, and residual distance modulus are retained.

\begin{table}[tbh]
\caption{Description of the columns included in the printed photometric AGN golden sample catalog.}
\label{tab:golden_catalog_columns}
\centering
\begin{tabular}{ll}
\hline\hline
Column & Description\\
\hline
objID & SDSS object identifier\\
RA$_{\rm deg}$ & Right Ascension (degrees)\\
DEC$_{\rm deg}$ & Declination (degrees)\\
$z$ & Spectroscopic redshift\\
$z_{\rm bin}$ & Redshift interval adopted in the pooling analysis\\
$D_{\rm phot}^{2}$ & Multidimensional statistical distance\\
$\mu_{\rm cand}$ & Candidate distance modulus after posterior projection\\
$\mu_{\rm proj}^{\rm post}$ & Posterior-calibrated distance modulus\\
$\Delta\mu_{\rm post}$ & Residual $\mu$ relative to the fiducial reconstruction\\
\hline
\end{tabular}
\end{table}

\begin{landscape}
\begingroup
\scriptsize
\setlength{\tabcolsep}{2.0pt}
\renewcommand{\arraystretch}{1.0}
\setlength{\LTcapwidth}{0.96\linewidth}
\setlength{\LTleft}{\fill}
\setlength{\LTright}{\fill}

\begin{longtable}{@{}lrrrrcrrr@{}}
\caption{Complete photometric AGN golden sample. Column definitions are given in Table~\ref{tab:golden_catalog_columns}.}
\label{tab:golden_sample_complete}\\
\hline\hline
objID &
RA$_{\rm deg}$ &
DEC$_{\rm deg}$ &
$z$ &
$z_{\rm bin}$ &
$D_{\rm phot}^{2}$ &
$\mu_{\rm cand}$ &
$\mu_{\rm proj}^{\rm post}$ &
$\Delta\mu_{\rm post}$\\
\hline
\endfirsthead

\caption{continued.}\\
\hline\hline
objID &
RA$_{\rm deg}$ &
DEC$_{\rm deg}$ &
$z$ &
$z_{\rm bin}$ &
$D_{\rm phot}^{2}$ &
$\mu_{\rm cand}$ &
$\mu_{\rm proj}^{\rm post}$ &
$\Delta\mu_{\rm post}$\\
\hline
\endhead

\hline
\endfoot

\hline
\endlastfoot

\multicolumn{9}{l}{\textbf{Redshift interval $0.0\leq z < 0.5$}}\\
\hline
1237651067893187208 & 202.22148 & 65.10646 & 0.09370 & 0.0--0.5 & 0.08954 & 38.10813 & 38.09819 & 0.00993\\
1237648703521030325 & 223.76828 & -0.65597 & 0.41888 & 0.0--0.5 & 0.22468 & 41.73285 & 41.73561 & -0.00277\\
1237648720670753572 & 127.28207 & -0.57441 & 0.42215 & 0.0--0.5 & 0.05683 & 41.76842 & 41.75547 & 0.01295\\
\multicolumn{9}{l}{\textbf{Redshift interval $0.5\leq z < 1.0$}}\\
\hline
1237648705655472296 & 193.99815 & 0.85168 & 0.50047 & 0.5--1.0 & 0.29846 & 42.19044 & 42.19512 & -0.00468\\
1237648705666613550 & 219.50827 & 0.97254 & 0.57001 & 0.5--1.0 & 0.36066 & 42.52950 & 42.53418 & -0.00468\\
1237648703522734728 & 227.70053 & -0.69969 & 0.58509 & 0.5--1.0 & 0.27732 & 42.61781 & 42.60246 & 0.01535\\
1237646586632929716 & 72.94804 & -0.15951 & 0.59432 & 0.5--1.0 & 0.21979 & 42.63408 & 42.64346 & -0.00939\\
1237648705651474671 & 184.91359 & 0.91658 & 0.60989 & 0.5--1.0 & 0.34430 & 42.72226 & 42.71141 & 0.01085\\
1237648705651867853 & 185.83777 & 0.96270 & 0.61702 & 0.5--1.0 & 0.12960 & 42.73258 & 42.74185 & -0.00927\\
1237648720152166728 & 168.94897 & -0.96964 & 0.62034 & 0.5--1.0 & 0.24578 & 42.74674 & 42.75605 & -0.00931\\
1237648704041779599 & 186.99300 & -0.25551 & 0.63050 & 0.5--1.0 & 0.26996 & 42.80355 & 42.79871 & 0.00484\\
1237648704597393747 & 229.80987 & 0.19396 & 0.64178 & 0.5--1.0 & 0.36222 & 42.83434 & 42.84559 & -0.01125\\
1237648705656324330 & 195.90487 & 0.86610 & 0.65514 & 0.5--1.0 & 0.22658 & 42.90604 & 42.89976 & 0.00628\\
1237648720141353336 & 144.30115 & -1.00719 & 0.65596 & 0.5--1.0 & 0.22765 & 42.88494 & 42.90310 & -0.01816\\
1237648720156295556 & 178.47487 & -0.92527 & 0.67822 & 0.5--1.0 & 0.12970 & 42.98669 & 42.99121 & -0.00452\\
1237648702983242162 & 221.66664 & -1.15638 & 0.68086 & 0.5--1.0 & 0.06534 & 43.00274 & 43.00139 & 0.00135\\
1237648702975574469 & 204.25962 & -1.22458 & 0.70072 & 0.5--1.0 & 0.36993 & 43.06643 & 43.07732 & -0.01089\\
1237648705668251983 & 223.27550 & 0.93906 & 0.71567 & 0.5--1.0 & 0.06497 & 43.13547 & 43.13325 & 0.00222\\
1237648705658749220 & 201.50160 & 0.85143 & 0.75165 & 0.5--1.0 & 0.06029 & 43.27074 & 43.26323 & 0.00751\\
1237646647297311053 & 71.57259 & 0.05056 & 0.75464 & 0.5--1.0 & 0.09488 & 43.26168 & 43.27370 & -0.01201\\
1237648720165732461 & 199.90197 & -1.00069 & 0.75481 & 0.5--1.0 & 0.16542 & 43.26847 & 43.27430 & -0.00583\\
1237648702972887236 & 198.11100 & -1.05292 & 0.75516 & 0.5--1.0 & 0.03857 & 43.26349 & 43.27552 & -0.01203\\
1237648704598770095 & 232.90545 & 0.16208 & 0.75814 & 0.5--1.0 & 0.17400 & 43.28572 & 43.28599 & -0.00027\\
1237648705131970757 & 224.49476 & 0.61100 & 0.78970 & 0.5--1.0 & 0.39352 & 43.41072 & 43.39427 & 0.01645\\
1237648702985142805 & 226.09587 & -1.08324 & 0.81888 & 0.5--1.0 & 0.24283 & 43.49502 & 43.49084 & 0.00418\\
1237648703523979840 & 230.60217 & -0.65629 & 0.86946 & 0.5--1.0 & 0.24647 & 43.65441 & 43.65047 & 0.00394\\
1237648720137355531 & 135.14695 & -0.93826 & 0.88350 & 0.5--1.0 & 0.37603 & 43.70787 & 43.69310 & 0.01477\\
1237648720137748854 & 136.05474 & -1.04932 & 0.88403 & 0.5--1.0 & 0.21746 & 43.69811 & 43.69469 & 0.00343\\
1237648720681173404 & 151.03551 & -0.46557 & 0.91005 & 0.5--1.0 & 0.31053 & 43.76971 & 43.77202 & -0.00231\\
1237648721208410824 & 129.06636 & -0.11747 & 0.93056 & 0.5--1.0 & 0.25051 & 43.82543 & 43.83145 & -0.00602\\
1237646586632864080 & 72.74386 & -0.18406 & 0.95953 & 0.5--1.0 & 0.33224 & 43.89895 & 43.91335 & -0.01440\\
1237648705666285743 & 218.64871 & 0.87084 & 0.96524 & 0.5--1.0 & 0.33828 & 43.91994 & 43.92916 & -0.00921\\
\multicolumn{9}{l}{\textbf{Redshift interval $1.0\leq z < 1.5$}}\\
\hline
1237646586100122512 & 82.28528 & -0.45536 & 1.00740 & 1.0--1.5 & 0.37728 & 44.05026 & 44.04339 & 0.00687\\
1237648704598377022 & 232.03033 & 0.01052 & 1.04262 & 1.0--1.5 & 0.22225 & 44.14172 & 44.13524 & 0.00648\\
1237648705651015952 & 183.85473 & 0.88804 & 1.04972 & 1.0--1.5 & 0.30439 & 44.14565 & 44.15335 & -0.00770\\
1237648702966857970 & 184.31273 & -1.05573 & 1.05151 & 1.0--1.5 & 0.23843 & 44.14435 & 44.15789 & -0.01354\\
1237646798138180456 & 122.58897 & 0.96576 & 1.07081 & 1.0--1.5 & 0.02305 & 44.21874 & 44.20650 & 0.01223\\
1237648705653244069 & 188.88634 & 0.91887 & 1.09583 & 1.0--1.5 & 0.32483 & 44.28695 & 44.26832 & 0.01863\\
1237648674530066977 & 239.07815 & 0.23703 & 1.13882 & 1.0--1.5 & 0.20812 & 44.35441 & 44.37119 & -0.01678\\
1237648703523127844 & 228.54196 & -0.74262 & 1.14857 & 1.0--1.5 & 0.31881 & 44.38785 & 44.39403 & -0.00618\\
1237648705655144725 & 193.32924 & 0.88610 & 1.18416 & 1.0--1.5 & 0.28085 & 44.48228 & 44.47566 & 0.00663\\
1237648705652916838 & 188.24578 & 0.84731 & 1.23543 & 1.0--1.5 & 0.38489 & 44.60849 & 44.58898 & 0.01951\\
1237645943978328209 & 54.98833 & 0.92120 & 1.24229 & 1.0--1.5 & 0.33711 & 44.59419 & 44.60377 & -0.00958\\
1237648703510675877 & 200.23101 & -0.83234 & 1.28326 & 1.0--1.5 & 0.16849 & 44.68680 & 44.69051 & -0.00371\\
1237648705123778736 & 205.81910 & 0.62277 & 1.34772 & 1.0--1.5 & 0.35172 & 44.83777 & 44.82152 & 0.01625\\
1237648705666875803 & 220.09190 & 1.01746 & 1.37432 & 1.0--1.5 & 0.17981 & 44.87231 & 44.87375 & -0.00144\\
1237648720134996621 & 129.78978 & -1.01871 & 1.37729 & 1.0--1.5 & 0.14801 & 44.88677 & 44.87954 & 0.00723\\
1237648705131119120 & 222.66540 & 0.55112 & 1.41845 & 1.0--1.5 & 0.19841 & 44.94525 & 44.95816 & -0.01290\\
1237648721205724137 & 122.86728 & -0.18961 & 1.42190 & 1.0--1.5 & 0.40750 & 44.97186 & 44.96464 & 0.00723\\
1237648720676978811 & 141.40817 & -0.61831 & 1.43102 & 1.0--1.5 & 0.39594 & 44.98104 & 44.98175 & -0.00072\\
1237648703518474739 & 217.89764 & -0.70881 & 1.44155 & 1.0--1.5 & 0.09528 & 45.01772 & 45.00128 & 0.01644\\
1237648721213260129 & 140.05670 & -0.07410 & 1.45037 & 1.0--1.5 & 0.02731 & 45.01934 & 45.01760 & 0.00174\\
\multicolumn{9}{l}{\textbf{Redshift interval $1.5\leq z < 2.0$}}\\
\hline
1237648705651540176 & 184.97422 & 0.99735 & 1.50727 & 1.5--2.0 & 0.27436 & 45.11428 & 45.12026 & -0.00598\\
1237648705130725794 & 221.75922 & 0.59078 & 1.51895 & 1.5--2.0 & 0.05699 & 45.14272 & 45.14088 & 0.00184\\
1237646648370725226 & 70.86013 & 0.87734 & 1.61838 & 1.5--2.0 & 0.35831 & 45.30546 & 45.30985 & -0.00439\\
1237648705652588816 & 187.46116 & 0.98089 & 1.64070 & 1.5--2.0 & 0.15773 & 45.35350 & 45.34636 & 0.00714\\
1237645943978983610 & 56.53608 & 0.88975 & 1.72036 & 1.5--2.0 & 0.34064 & 45.46212 & 45.47249 & -0.01037\\
1237648703524897186 & 232.67451 & -0.69568 & 1.72273 & 1.5--2.0 & 0.27982 & 45.45998 & 45.47615 & -0.01617\\
1237646647297966422 & 73.04501 & 0.14632 & 1.72646 & 1.5--2.0 & 0.26611 & 45.47189 & 45.48193 & -0.01004\\
1237648720147841477 & 159.15658 & -0.85477 & 1.76595 & 1.5--2.0 & 0.24558 & 45.56036 & 45.54205 & 0.01831\\
1237646645686632889 & 71.41574 & -1.15045 & 1.79067 & 1.5--2.0 & 0.31424 & 45.58553 & 45.57896 & 0.00658\\
1237648705656455396 & 196.20945 & 0.87478 & 1.81488 & 1.5--2.0 & 0.14820 & 45.60405 & 45.61465 & -0.01060\\
1237648704578781447 & 187.28180 & 0.17923 & 1.82127 & 1.5--2.0 & 0.19650 & 45.63422 & 45.62396 & 0.01027\\
1237648720678289485 & 144.43786 & -0.55895 & 1.84787 & 1.5--2.0 & 0.05512 & 45.66439 & 45.66245 & 0.00194\\
1237648704049774750 & 205.24685 & -0.32915 & 1.85944 & 1.5--2.0 & 0.34567 & 45.69838 & 45.67901 & 0.01937\\
1237648721236525374 & 193.22723 & -0.09882 & 1.86599 & 1.5--2.0 & 0.34257 & 45.67431 & 45.68835 & -0.01404\\
1237648721755177226 & 151.65829 & 0.24667 & 1.91322 & 1.5--2.0 & 0.10582 & 45.74808 & 45.75463 & -0.00655\\
1237648720674292095 & 135.35765 & -0.61754 & 1.91583 & 1.5--2.0 & 0.05276 & 45.75040 & 45.75824 & -0.00784\\
1237648721750589610 & 141.15232 & 0.41952 & 1.96340 & 1.5--2.0 & 0.16715 & 45.82916 & 45.82324 & 0.00592\\
1237645943978655930 & 55.71857 & 0.94825 & 1.96587 & 1.5--2.0 & 0.23131 & 45.83178 & 45.82656 & 0.00522\\
1237648720676258185 & 139.80291 & -0.49502 & 1.96628 & 1.5--2.0 & 0.06796 & 45.82722 & 45.82712 & 0.00010\\
1237648721753342335 & 147.44136 & 0.40758 & 1.98474 & 1.5--2.0 & 0.14308 & 45.84717 & 45.85186 & -0.00469\\
\multicolumn{9}{l}{\textbf{Redshift interval $2.0\leq z < 2.5$}}\\
\hline
1237648703520965191 & 223.68899 & -0.67324 & 2.00828 & 2.0--2.5 & 0.14860 & 45.87467 & 45.88309 & -0.00841\\
1237648721211556315 & 136.26682 & -0.07894 & 2.01107 & 2.0--2.5 & 0.14052 & 45.87991 & 45.88676 & -0.00685\\
1237648703521227276 & 224.28866 & -0.65190 & 2.01395 & 2.0--2.5 & 0.34682 & 45.87449 & 45.89054 & -0.01605\\
1237648720678879352 & 145.72396 & -0.61476 & 2.02113 & 2.0--2.5 & 0.16035 & 45.89672 & 45.89995 & -0.00323\\
1237648721209131611 & 130.65359 & -0.16087 & 2.02540 & 2.0--2.5 & 0.07954 & 45.91363 & 45.90555 & 0.00808\\
1237648721213718791 & 141.19414 & -0.06731 & 2.03354 & 2.0--2.5 & 0.23544 & 45.91964 & 45.91614 & 0.00349\\
1237648702968365338 & 187.68941 & -1.25111 & 2.03962 & 2.0--2.5 & 0.29273 & 45.92920 & 45.92405 & 0.00515\\
1237648704583958912 & 199.13920 & 0.18666 & 2.04645 & 2.0--2.5 & 0.16477 & 45.93141 & 45.93290 & -0.00149\\
1237646587169931802 & 73.30160 & 0.39015 & 2.07404 & 2.0--2.5 & 0.17203 & 45.96778 & 45.96831 & -0.00053\\
1237645943978459453 & 55.33151 & 1.02675 & 2.07548 & 2.0--2.5 & 0.30040 & 45.97135 & 45.97014 & 0.00120\\
1237648704599097428 & 233.66461 & 0.05512 & 2.08857 & 2.0--2.5 & 0.21015 & 45.98291 & 45.98676 & -0.00385\\
1237648705130987808 & 222.30601 & 0.58413 & 2.09230 & 2.0--2.5 & 0.17665 & 45.99751 & 45.99148 & 0.00603\\
1237648674510864617 & 195.36333 & 0.31364 & 2.11211 & 2.0--2.5 & 0.39166 & 46.01568 & 46.01638 & -0.00069\\
1237648704049971398 & 205.63729 & -0.26407 & 2.13403 & 2.0--2.5 & 0.40475 & 46.05869 & 46.04365 & 0.01504\\
1237648704579567926 & 189.01300 & 0.05962 & 2.15409 & 2.0--2.5 & 0.20928 & 46.08417 & 46.06835 & 0.01582\\
1237648702969086159 & 189.30796 & -1.15213 & 2.16340 & 2.0--2.5 & 0.26391 & 46.06155 & 46.07974 & -0.01819\\
1237648702977868083 & 209.38842 & -1.06349 & 2.17774 & 2.0--2.5 & 0.11341 & 46.10890 & 46.09717 & 0.01174\\
1237648704599622207 & 234.90001 & 0.16496 & 2.18279 & 2.0--2.5 & 0.25243 & 46.08348 & 46.10329 & -0.01981\\
1237648704597656000 & 230.38856 & 0.16180 & 2.18444 & 2.0--2.5 & 0.35401 & 46.09397 & 46.10530 & -0.01133\\
1237648703512379562 & 204.10730 & -0.77511 & 2.19058 & 2.0--2.5 & 0.38581 & 46.11468 & 46.11269 & 0.00199\\
1237648705122271374 & 202.38063 & 0.42563 & 2.19403 & 2.0--2.5 & 0.20127 & 46.10676 & 46.11683 & -0.01007\\
1237648703505957027 & 189.36665 & -0.73704 & 2.19529 & 2.0--2.5 & 0.24510 & 46.11972 & 46.11835 & 0.00137\\
1237648705115652233 & 187.33371 & 0.48520 & 2.19598 & 2.0--2.5 & 0.33797 & 46.12865 & 46.11918 & 0.00947\\
1237648704047939976 & 201.05363 & -0.36140 & 2.19631 & 2.0--2.5 & 0.33556 & 46.10119 & 46.11958 & -0.01839\\
1237648705124303316 & 207.08301 & 0.43261 & 2.19810 & 2.0--2.5 & 0.15683 & 46.10606 & 46.12173 & -0.01567\\
1237648702978392413 & 210.62261 & -1.08678 & 2.21059 & 2.0--2.5 & 0.37748 & 46.13022 & 46.13666 & -0.00644\\
1237648703515066771 & 210.16032 & -0.73902 & 2.22148 & 2.0--2.5 & 0.20134 & 46.16592 & 46.14965 & 0.01627\\
1237646798138573844 & 123.53638 & 0.98210 & 2.22328 & 2.0--2.5 & 0.33276 & 46.15632 & 46.15177 & 0.00455\\
1237646586632601984 & 72.20765 & -0.07775 & 2.22866 & 2.0--2.5 & 0.25606 & 46.16183 & 46.15813 & 0.00369\\
1237648702968823923 & 188.79625 & -1.14153 & 2.23186 & 2.0--2.5 & 0.23209 & 46.16712 & 46.16192 & 0.00520\\
1237648703514214668 & 208.30437 & -0.70187 & 2.23599 & 2.0--2.5 & 0.23999 & 46.18245 & 46.16681 & 0.01564\\
1237646647297179952 & 71.28339 & 0.18105 & 2.24243 & 2.0--2.5 & 0.29990 & 46.18482 & 46.17438 & 0.01044\\
1237648674511847634 & 197.48959 & 0.21892 & 2.24301 & 2.0--2.5 & 0.12911 & 46.18100 & 46.17506 & 0.00594\\
1237648703522079238 & 226.17041 & -0.76764 & 2.24610 & 2.0--2.5 & 0.32312 & 46.18370 & 46.17869 & 0.00501\\
1237648702978195775 & 210.11076 & -1.21971 & 2.25685 & 2.0--2.5 & 0.03609 & 46.19821 & 46.19129 & 0.00693\\
1237648702971445613 & 194.83836 & -1.09013 & 2.26383 & 2.0--2.5 & 0.38445 & 46.18581 & 46.19941 & -0.01360\\
1237648703512576321 & 204.43890 & -0.73035 & 2.27844 & 2.0--2.5 & 0.40823 & 46.22100 & 46.21636 & 0.00465\\
1237648702978457996 & 210.72758 & -1.15254 & 2.28027 & 2.0--2.5 & 0.32452 & 46.22433 & 46.21847 & 0.00586\\
1237648704053182591 & 212.97350 & -0.35486 & 2.28042 & 2.0--2.5 & 0.18455 & 46.20436 & 46.21864 & -0.01429\\
1237648704063341272 & 236.27207 & -0.35471 & 2.29341 & 2.0--2.5 & 0.33515 & 46.24309 & 46.23362 & 0.00948\\
1237648704040599798 & 184.21568 & -0.28867 & 2.29971 & 2.0--2.5 & 0.39744 & 46.25396 & 46.24083 & 0.01313\\
1237648703516377501 & 213.21624 & -0.79263 & 2.30533 & 2.0--2.5 & 0.20468 & 46.22778 & 46.24728 & -0.01949\\
1237648704043221226 & 190.21009 & -0.29530 & 2.30591 & 2.0--2.5 & 0.12771 & 46.26367 & 46.24793 & 0.01574\\
1237648703511331170 & 201.70469 & -0.68481 & 2.31272 & 2.0--2.5 & 0.25245 & 46.23928 & 46.25569 & -0.01641\\
1237646647297835519 & 72.84754 & 0.04635 & 2.31632 & 2.0--2.5 & 0.21005 & 46.27905 & 46.25978 & 0.01927\\
1237648704046629028 & 198.09875 & -0.21085 & 2.32800 & 2.0--2.5 & 0.28926 & 46.28129 & 46.27303 & 0.00826\\
1237648703506219147 & 189.96022 & -0.74107 & 2.33087 & 2.0--2.5 & 0.12278 & 46.29271 & 46.27627 & 0.01644\\
1237648703518409013 & 217.86845 & -0.74724 & 2.33928 & 2.0--2.5 & 0.21127 & 46.28085 & 46.28575 & -0.00491\\
1237648704043024726 & 189.83412 & -0.36260 & 2.35849 & 2.0--2.5 & 0.37850 & 46.29220 & 46.30728 & -0.01507\\
1237648704068715146 & 248.49115 & -0.33680 & 2.36889 & 2.0--2.5 & 0.16586 & 46.30105 & 46.31884 & -0.01779\\
1237648704044007653 & 192.04919 & -0.37257 & 2.37352 & 2.0--2.5 & 0.38547 & 46.30766 & 46.32398 & -0.01632\\
1237648702982324520 & 219.63650 & -1.14760 & 2.37883 & 2.0--2.5 & 0.12005 & 46.32832 & 46.32986 & -0.00154\\
1237648704061964559 & 233.03530 & -0.38653 & 2.38844 & 2.0--2.5 & 0.40608 & 46.34728 & 46.34046 & 0.00682\\
1237646647833985378 & 71.17379 & 0.49113 & 2.39362 & 2.0--2.5 & 0.23575 & 46.35169 & 46.34616 & 0.00553\\
1237646586632143422 & 71.23719 & -0.13597 & 2.39531 & 2.0--2.5 & 0.30786 & 46.36456 & 46.34802 & 0.01654\\
1237648704049969920 & 205.75529 & -0.23602 & 2.40023 & 2.0--2.5 & 0.16556 & 46.35366 & 46.35340 & 0.00026\\
1237648703504646289 & 186.39948 & -0.77726 & 2.41190 & 2.0--2.5 & 0.33544 & 46.36087 & 46.36617 & -0.00529\\
1237648703507202221 & 192.23624 & -0.69053 & 2.41514 & 2.0--2.5 & 0.08241 & 46.38220 & 46.36968 & 0.01251\\
1237646647297769992 & 72.59366 & 0.08084 & 2.41813 & 2.0--2.5 & 0.05206 & 46.35579 & 46.37293 & -0.01714\\
1237648704581533950 & 193.52712 & 0.15125 & 2.42117 & 2.0--2.5 & 0.06464 & 46.38588 & 46.37623 & 0.00965\\
1237648703505891615 & 189.25847 & -0.63236 & 2.42197 & 2.0--2.5 & 0.32269 & 46.37263 & 46.37710 & -0.00447\\
1237648703522734697 & 227.66160 & -0.75810 & 2.42883 & 2.0--2.5 & 0.34387 & 46.39882 & 46.38454 & 0.01428\\
1237648702971380102 & 194.64762 & -1.22208 & 2.43636 & 2.0--2.5 & 0.08937 & 46.40213 & 46.39267 & 0.00946\\
1237648703508644202 & 195.44547 & -0.82408 & 2.44066 & 2.0--2.5 & 0.30432 & 46.39584 & 46.39730 & -0.00146\\
1237648702968692957 & 188.48182 & -1.20107 & 2.47621 & 2.0--2.5 & 0.33524 & 46.41835 & 46.43528 & -0.01694\\
1237648703521030663 & 223.86625 & -0.71725 & 2.48018 & 2.0--2.5 & 0.19198 & 46.43056 & 46.43949 & -0.00893\\
1237648704045711686 & 196.00624 & -0.27317 & 2.48891 & 2.0--2.5 & 0.40522 & 46.46584 & 46.44872 & 0.01713\\
1237648702977343744 & 208.30376 & -1.07058 & 2.49351 & 2.0--2.5 & 0.27434 & 46.45783 & 46.45355 & 0.00428\\
\multicolumn{9}{l}{\textbf{Redshift interval $2.5\leq z < 3.0$}}\\
\hline
1237648705129808183 & 219.61998 & 0.60835 & 2.50485 & 2.5--3.0 & 0.19261 & 46.46893 & 46.46547 & 0.00346\\
1237648705136951666 & 235.91008 & 0.46238 & 2.50799 & 2.5--3.0 & 0.35514 & 46.47646 & 46.46875 & 0.00771\\
1237646794375430893 & 81.65141 & 1.39196 & 2.50937 & 2.5--3.0 & 0.23140 & 46.45252 & 46.47020 & -0.01768\\
1237648704578650258 & 187.05410 & 0.18240 & 2.52211 & 2.5--3.0 & 0.18521 & 46.48657 & 46.48349 & 0.00307\\
1237646587169669645 & 72.71011 & 0.30023 & 2.52494 & 2.5--3.0 & 0.39133 & 46.48759 & 46.48643 & 0.00116\\
1237648704063013354 & 235.52871 & -0.22127 & 2.52911 & 2.5--3.0 & 0.10906 & 46.49792 & 46.49076 & 0.00716\\
1237648705120305307 & 197.88603 & 0.61253 & 2.53313 & 2.5--3.0 & 0.40542 & 46.50615 & 46.49493 & 0.01122\\
1237648704054427853 & 215.88104 & -0.36806 & 2.56204 & 2.5--3.0 & 0.21202 & 46.50593 & 46.52468 & -0.01876\\
1237648705135182404 & 231.92779 & 0.51300 & 2.57307 & 2.5--3.0 & 0.24281 & 46.54625 & 46.53595 & 0.01030\\
1237648703517950307 & 216.79153 & -0.74743 & 2.58170 & 2.5--3.0 & 0.39957 & 46.55500 & 46.54473 & 0.01028\\
1237648705119125731 & 195.15850 & 0.57300 & 2.58444 & 2.5--3.0 & 0.18276 & 46.55781 & 46.54751 & 0.01030\\
1237648704058229263 & 224.49400 & -0.23185 & 2.60341 & 2.5--3.0 & 0.30392 & 46.57810 & 46.56668 & 0.01142\\
1237648704577208453 & 183.62680 & 0.15404 & 2.61110 & 2.5--3.0 & 0.26048 & 46.59348 & 46.57440 & 0.01908\\
1237646648371380554 & 72.33904 & 0.96830 & 2.62247 & 2.5--3.0 & 0.22053 & 46.59881 & 46.58579 & 0.01303\\
1237648705115390170 & 186.62433 & 0.48163 & 2.63941 & 2.5--3.0 & 0.16766 & 46.59634 & 46.60265 & -0.00631\\
1237646646761160926 & 73.18555 & -0.25224 & 2.64110 & 2.5--3.0 & 0.26377 & 46.61951 & 46.60433 & 0.01518\\
1237648704596869473 & 228.63941 & 0.03034 & 2.65192 & 2.5--3.0 & 0.14655 & 46.59957 & 46.61502 & -0.01546\\
1237648703520899348 & 223.46457 & -0.68366 & 2.65217 & 2.5--3.0 & 0.32598 & 46.61053 & 46.61526 & -0.00474\\
1237648704040199936 & 183.37133 & -0.31704 & 2.65602 & 2.5--3.0 & 0.18743 & 46.60936 & 46.61907 & -0.00971\\
1237648705128169807 & 215.88435 & 0.45317 & 2.65821 & 2.5--3.0 & 0.26371 & 46.62287 & 46.62124 & 0.00164\\
1237646647297966563 & 73.15760 & 0.02683 & 2.66895 & 2.5--3.0 & 0.24855 & 46.64550 & 46.63178 & 0.01373\\
1237648705656848527 & 197.15604 & 1.02446 & 2.68308 & 2.5--3.0 & 0.32592 & 46.63615 & 46.64559 & -0.00944\\
1237648705136558687 & 235.03989 & 0.62329 & 2.70488 & 2.5--3.0 & 0.07941 & 46.66806 & 46.66677 & 0.00129\\
1237648720672915793 & 132.16230 & -0.55785 & 2.71779 & 2.5--3.0 & 0.12464 & 46.66492 & 46.67922 & -0.01431\\
1237648704589856969 & 212.64140 & 0.10747 & 2.72588 & 2.5--3.0 & 0.29016 & 46.70060 & 46.68701 & 0.01359\\
1237648720146334032 & 155.69708 & -1.04318 & 2.72967 & 2.5--3.0 & 0.11227 & 46.67603 & 46.69064 & -0.01461\\
1237648720141287782 & 144.17574 & -0.97536 & 2.73043 & 2.5--3.0 & 0.31071 & 46.69570 & 46.69137 & 0.00433\\
1237648721207492800 & 126.91475 & -0.12775 & 2.73059 & 2.5--3.0 & 0.16828 & 46.69243 & 46.69152 & 0.00091\\
1237648720141811744 & 145.27540 & -0.86758 & 2.74436 & 2.5--3.0 & 0.30257 & 46.69164 & 46.70467 & -0.01303\\
1237648720712892753 & 223.48412 & -0.60149 & 2.75672 & 2.5--3.0 & 0.39323 & 46.72909 & 46.71641 & 0.01268\\
1237648720677831221 & 143.38125 & -0.46825 & 2.77811 & 2.5--3.0 & 0.16884 & 46.74034 & 46.73661 & 0.00373\\
1237648704051282168 & 208.74943 & -0.34123 & 2.78281 & 2.5--3.0 & 0.37979 & 46.72779 & 46.74103 & -0.01324\\
1237648704063078772 & 235.63486 & -0.34801 & 2.80192 & 2.5--3.0 & 0.26315 & 46.77595 & 46.75890 & 0.01705\\
1237648721219355054 & 154.08463 & -0.02683 & 2.80339 & 2.5--3.0 & 0.38440 & 46.74442 & 46.76027 & -0.01585\\
1237648720159113590 & 184.88891 & -0.93400 & 2.80527 & 2.5--3.0 & 0.26313 & 46.75973 & 46.76202 & -0.00229\\
1237648704595100258 & 224.61469 & 0.02795 & 2.80652 & 2.5--3.0 & 0.18298 & 46.77328 & 46.76318 & 0.01010\\
1237648721223286985 & 163.02836 & -0.14532 & 2.80826 & 2.5--3.0 & 0.24519 & 46.78211 & 46.76480 & 0.01732\\
1237648702979440817 & 212.97086 & -1.21334 & 2.82122 & 2.5--3.0 & 0.12158 & 46.77014 & 46.77683 & -0.00669\\
1237648720687071969 & 164.53350 & -0.49498 & 2.82420 & 2.5--3.0 & 0.31647 & 46.77008 & 46.77958 & -0.00950\\
1237648720693821953 & 179.94083 & -0.61307 & 2.83067 & 2.5--3.0 & 0.09042 & 46.77249 & 46.78556 & -0.01307\\
1237648720673374821 & 133.24775 & -0.44716 & 2.84223 & 2.5--3.0 & 0.12214 & 46.80429 & 46.79619 & 0.00810\\
1237648720169337019 & 208.17912 & -0.90675 & 2.86257 & 2.5--3.0 & 0.32669 & 46.83316 & 46.81481 & 0.01836\\
1237648704592020012 & 217.45140 & 0.20413 & 2.87800 & 2.5--3.0 & 0.10822 & 46.81915 & 46.82883 & -0.00968\\
1237648721237246472 & 194.84422 & -0.20279 & 2.88382 & 2.5--3.0 & 0.08129 & 46.81995 & 46.83410 & -0.01415\\
1237648720143843576 & 149.90053 & -0.86867 & 2.88694 & 2.5--3.0 & 0.39815 & 46.84581 & 46.83694 & 0.00888\\
1237648703524110817 & 230.79604 & -0.75604 & 2.90487 & 2.5--3.0 & 0.19722 & 46.86857 & 46.85308 & 0.01549\\
1237648704047547015 & 200.16623 & -0.26155 & 2.91175 & 2.5--3.0 & 0.05557 & 46.84070 & 46.85924 & -0.01854\\
1237648720143974606 & 150.20642 & -0.85508 & 2.94330 & 2.5--3.0 & 0.35760 & 46.88614 & 46.88733 & -0.00119\\
1237648702993858575 & 245.89531 & -1.07825 & 2.94997 & 2.5--3.0 & 0.38192 & 46.88818 & 46.89323 & -0.00505\\
\multicolumn{9}{l}{\textbf{Redshift interval $3.0\leq z < 3.5$}}\\
\hline
1237648704057770234 & 223.46303 & -0.29904 & 3.01572 & 3.0--3.5 & 0.24505 & 46.94184 & 46.95064 & -0.00880\\
1237648703506088450 & 189.67186 & -0.66557 & 3.02409 & 3.0--3.5 & 0.23539 & 46.96048 & 46.95787 & 0.00261\\
1237648705667334376 & 221.13610 & 0.88900 & 3.03281 & 3.0--3.5 & 0.21302 & 46.96556 & 46.96536 & 0.00020\\
1237648721232527464 & 184.03415 & -0.09271 & 3.03588 & 3.0--3.5 & 0.39657 & 46.94836 & 46.96799 & -0.01963\\
1237648705115783636 & 187.55230 & 0.46529 & 3.03955 & 3.0--3.5 & 0.24679 & 46.98037 & 46.97113 & 0.00924\\
1237648720150004200 & 164.10802 & -0.86215 & 3.04037 & 3.0--3.5 & 0.40235 & 46.97582 & 46.97184 & 0.00398\\
1237648720142664161 & 147.29070 & -0.90444 & 3.04409 & 3.0--3.5 & 0.24419 & 46.96302 & 46.97502 & -0.01200\\
1237648704058294849 & 224.72524 & -0.21712 & 3.04483 & 3.0--3.5 & 0.09440 & 46.97705 & 46.97566 & 0.00140\\
1237648721232724816 & 184.59472 & -0.00334 & 3.06666 & 3.0--3.5 & 0.38514 & 46.98654 & 46.99424 & -0.00771\\
1237648704045449467 & 195.30112 & -0.34694 & 3.08637 & 3.0--3.5 & 0.25076 & 47.01317 & 47.01090 & 0.00227\\
1237646646760637331 & 72.06406 & -0.36215 & 3.08843 & 3.0--3.5 & 0.20376 & 47.02588 & 47.01264 & 0.01324\\
1237648720700768855 & 195.80212 & -0.42652 & 3.12350 & 3.0--3.5 & 0.33343 & 47.03896 & 47.04199 & -0.00304\\
1237648704597656069 & 230.32311 & 0.18536 & 3.13008 & 3.0--3.5 & 0.09612 & 47.05924 & 47.04747 & 0.01177\\
1237648721214767476 & 143.57163 & -0.12019 & 3.13183 & 3.0--3.5 & 0.08525 & 47.05454 & 47.04892 & 0.00562\\
1237648704585597211 & 202.84091 & 0.04679 & 3.14703 & 3.0--3.5 & 0.20103 & 47.04904 & 47.06150 & -0.01245\\
1237648704585203884 & 201.96526 & 0.09164 & 3.15679 & 3.0--3.5 & 0.30276 & 47.06538 & 47.06954 & -0.00416\\
1237648702977278371 & 208.12963 & -1.13162 & 3.15867 & 3.0--3.5 & 0.31592 & 47.06123 & 47.07109 & -0.00986\\
1237648721210311071 & 133.28294 & -0.05351 & 3.16433 & 3.0--3.5 & 0.27282 & 47.06396 & 47.07575 & -0.01179\\
1237648702975640200 & 204.33428 & -1.17207 & 3.17147 & 3.0--3.5 & 0.27649 & 47.09858 & 47.08159 & 0.01699\\
1237648721253434237 & 231.87546 & -0.13338 & 3.17542 & 3.0--3.5 & 0.25947 & 47.09196 & 47.08482 & 0.00714\\
1237648720679534821 & 147.30681 & -0.43708 & 3.18044 & 3.0--3.5 & 0.39815 & 47.07507 & 47.08893 & -0.01386\\
1237648703523062582 & 228.40221 & -0.63799 & 3.18735 & 3.0--3.5 & 0.10808 & 47.11256 & 47.09457 & 0.01799\\
1237648720694673883 & 181.79466 & -0.59587 & 3.18817 & 3.0--3.5 & 0.09316 & 47.10138 & 47.09523 & 0.00615\\
1237648721759830446 & 162.28852 & 0.39164 & 3.19377 & 3.0--3.5 & 0.32820 & 47.11605 & 47.09979 & 0.01626\\
1237649920579928487 & 23.56711 & 14.81203 & 3.21359 & 3.0--3.5 & 0.19587 & 47.12844 & 47.11585 & 0.01259\\
1237648721765859865 & 175.96399 & 0.24129 & 3.21469 & 3.0--3.5 & 0.08389 & 47.10016 & 47.11674 & -0.01658\\
1237648720688382710 & 167.51571 & -0.50617 & 3.21873 & 3.0--3.5 & 0.04392 & 47.10881 & 47.12001 & -0.01120\\
1237648705132691576 & 226.25023 & 0.58825 & 3.24747 & 3.0--3.5 & 0.29199 & 47.15801 & 47.14307 & 0.01495\\
1237648721207231785 & 126.37784 & -0.13586 & 3.28332 & 3.0--3.5 & 0.37859 & 47.18375 & 47.17155 & 0.01221\\
1237648704040599985 & 184.25197 & -0.40399 & 3.29261 & 3.0--3.5 & 0.33377 & 47.17846 & 47.17888 & -0.00042\\
1237648721215160976 & 144.47454 & -0.05743 & 3.30165 & 3.0--3.5 & 0.16192 & 47.18833 & 47.18599 & 0.00235\\
1237650761856647635 & 186.73961 & -2.47452 & 3.31977 & 3.0--3.5 & 0.01863 & 47.18834 & 47.20017 & -0.01183\\
1237648704053706817 & 214.22188 & -0.35733 & 3.32660 & 3.0--3.5 & 0.26706 & 47.21644 & 47.20550 & 0.01094\\
1237651066280739233 & 191.87453 & 64.75940 & 3.32959 & 3.0--3.5 & 0.26780 & 47.20379 & 47.20783 & -0.00404\\
1237648722316493449 & 207.42187 & 0.66850 & 3.33072 & 3.0--3.5 & 0.30089 & 47.20435 & 47.20870 & -0.00435\\
1237648722837045840 & 170.21728 & 1.20272 & 3.33663 & 3.0--3.5 & 0.18424 & 47.22981 & 47.21330 & 0.01651\\
1237648721750327938 & 140.57093 & 0.34411 & 3.33701 & 3.0--3.5 & 0.23238 & 47.22986 & 47.21359 & 0.01627\\
1237650805882487132 & 161.43071 & -1.43029 & 3.34984 & 3.0--3.5 & 0.19175 & 47.22396 & 47.22354 & 0.00042\\
1237649918435328303 & 30.02361 & 12.45715 & 3.35512 & 3.0--3.5 & 0.34366 & 47.24126 & 47.22762 & 0.01364\\
1237650761315319919 & 176.53224 & -2.87336 & 3.35885 & 3.0--3.5 & 0.15658 & 47.22430 & 47.23051 & -0.00621\\
1237648705651933576 & 185.90384 & 1.04813 & 3.35991 & 3.0--3.5 & 0.23444 & 47.23780 & 47.23132 & 0.00648\\
1237650796215927595 & 136.84188 & 0.09242 & 3.36980 & 3.0--3.5 & 0.11440 & 47.23633 & 47.23893 & -0.00260\\
1237650795147166259 & 148.18788 & -0.75082 & 3.37074 & 3.0--3.5 & 0.09859 & 47.24159 & 47.23965 & 0.00194\\
1237650761857040736 & 187.61299 & -2.48534 & 3.37195 & 3.0--3.5 & 0.35473 & 47.25249 & 47.24058 & 0.01190\\
1237648704589398780 & 211.58475 & 0.14769 & 3.38281 & 3.0--3.5 & 0.14952 & 47.23254 & 47.24891 & -0.01636\\
1237650370478473647 & 172.05128 & -2.61718 & 3.40503 & 3.0--3.5 & 0.16491 & 47.28170 & 47.26586 & 0.01583\\
1237648722833703372 & 162.52842 & 1.13245 & 3.40871 & 3.0--3.5 & 0.40290 & 47.28114 & 47.26866 & 0.01248\\
1237648720140763654 & 142.89341 & -1.04909 & 3.41587 & 3.0--3.5 & 0.03907 & 47.28666 & 47.27409 & 0.01257\\
1237651191896277635 & 134.52928 & 54.43529 & 3.41977 & 3.0--3.5 & 0.22221 & 47.27680 & 47.27704 & -0.00024\\
1237651251506315779 & 220.85363 & 60.54793 & 3.42187 & 3.0--3.5 & 0.30435 & 47.26069 & 47.27863 & -0.01795\\
1237651191361831270 & 142.44176 & 57.77152 & 3.42618 & 3.0--3.5 & 0.33271 & 47.28341 & 47.28189 & 0.00152\\
1237651249896817107 & 223.50933 & 57.98219 & 3.42678 & 3.0--3.5 & 0.30306 & 47.26880 & 47.28235 & -0.01354\\
1237651190282912228 & 128.96671 & 49.13119 & 3.43198 & 3.0--3.5 & 0.38487 & 47.30577 & 47.28627 & 0.01950\\
1237648721247470416 & 218.28395 & -0.02911 & 3.43233 & 3.0--3.5 & 0.23230 & 47.28359 & 47.28653 & -0.00294\\
1237648703511659006 & 202.35653 & -0.68737 & 3.43838 & 3.0--3.5 & 0.21693 & 47.27458 & 47.29109 & -0.01651\\
1237650372093739889 & 182.78986 & -1.30357 & 3.44843 & 3.0--3.5 & 0.27311 & 47.29222 & 47.29865 & -0.00643\\
1237650370474344961 & 162.59171 & -2.38602 & 3.45104 & 3.0--3.5 & 0.34148 & 47.30886 & 47.30060 & 0.00826\\
1237649920575144353 & 12.19559 & 15.35035 & 3.46020 & 3.0--3.5 & 0.26477 & 47.29618 & 47.30746 & -0.01127\\
1237651252049215756 & 240.18290 & 51.74021 & 3.46069 & 3.0--3.5 & 0.20825 & 47.29138 & 47.30782 & -0.01644\\
1237651250410160851 & 127.26007 & 48.11696 & 3.46997 & 3.0--3.5 & 0.35566 & 47.31963 & 47.31475 & 0.00487\\
1237651274042376496 & 152.72139 & 65.09999 & 3.47216 & 3.0--3.5 & 0.28638 & 47.32128 & 47.31638 & 0.00490\\
1237648704044598124 & 193.48550 & -0.28629 & 3.47404 & 3.0--3.5 & 0.11575 & 47.30764 & 47.31778 & -0.01014\\
1237651252577698227 & 211.44409 & 63.77346 & 3.47620 & 3.0--3.5 & 0.13864 & 47.30997 & 47.31939 & -0.00942\\
1237651496832139448 & 119.85512 & 46.95664 & 3.47648 & 3.0--3.5 & 0.40501 & 47.32424 & 47.31960 & 0.00464\\
1237651190287827513 & 142.59396 & 56.77179 & 3.48487 & 3.0--3.5 & 0.24544 & 47.31147 & 47.32584 & -0.01437\\
1237650805882356153 & 161.18959 & -1.29903 & 3.49734 & 3.0--3.5 & 0.28138 & 47.33080 & 47.33507 & -0.00427\\
\end{longtable}
\endgroup
\end{landscape}



\begin{thebibliography}{}

\bibitem[Abbott et al.(2017)]{Abbott2017}
Abbott, B. P., Abbott, R., Abbott, T. D., et al. 2017,
Nature, 551, 85

\bibitem[Anand et al.(2022)]{Anand2022}
Anand, G. S., Tully, R. B., Rizzi, L., Riess, A. G., \& Yuan, W. 2022,
ApJ, 932, 15

\bibitem[Banados et al.(2018)]{Banados2018}
Bañados, E., Venemans, B. P., Mazzucchelli, C., et al. 2018,
Nature, 553, 473

\bibitem[Birrer et al.(2020)]{Birrer2020}
Birrer, S., Shajib, A. J., Galan, A., et al. 2020,
A\&A, 643, A165

\bibitem[Bisogni et al.(2021)]{Bisogni2021}
Bisogni, S., Lusso, E., Civano, F., et al. 2021,
A\&A, 655, A109

\bibitem[Blakeslee et al.(2021)]{Blakeslee2021}
Blakeslee, J. P., Jensen, J. B., Ma, C.-P., Milne, P. A., \& Greene, J. E. 2021,
ApJ, 911, 65

\bibitem[Caplar et al.(2015)]{Caplar2015}
Caplar, N., Lilly, S. J., \& Trakhtenbrot, B. 2015,
ApJ, 811, 148

\bibitem[Caplar et al.(2018)]{Caplar2018}
Caplar, N., Lilly, S. J., \& Trakhtenbrot, B. 2018,
ApJ, 867, 148

\bibitem[Ciesla et al.(2015)]{Ciesla2015}
Ciesla, L., Charmandaris, V., Georgakakis, A., et al. 2015,
A\&A, 576, A10

\bibitem[Di Valentino et al.(2021)]{DiValentino2021}
Di Valentino, E., Mena, O., Pan, S., et al. 2021,
Class. Quantum Grav., 38, 153001

\bibitem[Freedman et al.(2019)]{Freedman2019}
Freedman, W. L., Madore, B. F., Hatt, D., et al. 2019,
ApJ, 882, 34

\bibitem[Freedman(2021)]{Freedman2021}
Freedman, W. L. 2021,
ApJ, 919, 16

\bibitem[Fukugita et al.(1996)]{Fukugita1996}
Fukugita, M., Ichikawa, T., Gunn, J. E., et al. 1996,
AJ, 111, 1748

\bibitem[Garnavich et al.(2023)]{Garnavich2023}
Garnavich, P., Wood, C. M., Milne, P., et al. 2023,
ApJ, 953, 35

\bibitem[Gelman et al.(2013)]{Gelman2013}
Gelman, A., Carlin, J. B., Stern, H. S., Dunson, D. B.,
Vehtari, A., \& Rubin, D. B. 2013,
Bayesian Data Analysis, 3rd edn.
(Boca Raton, FL: CRC Press)

\bibitem[Haas et al.(2003)]{Haas2003}
Haas, M., Klaas, U., Müller, S. A. H., et al. 2003,
A\&A, 402, 87

\bibitem[Hogg(1999)]{Hogg1999}
Hogg, D. W. 1999,
arXiv:astro-ph/9905116

\bibitem[Huang et al.(2020)]{Huang2020}
Huang, C. D., Riess, A. G., Yuan, W., et al. 2020,
ApJ, 889, 5

\bibitem[Kelly et al.(2023)]{Kelly2023}
Kelly, P. L., Rodney, S., Treu, T., et al. 2023,
Science, 380, eabh1322

\bibitem[Lusso \& Risaliti(2016)]{Lusso2016}
Lusso, E., \& Risaliti, G. 2016,
ApJ, 819, 154

\bibitem[Lusso et al.(2020)]{Lusso2020}
Lusso, E., Risaliti, G., Nardini, E., et al. 2020,
A\&A, 642, A150

\bibitem[Mahalanobis(1936)]{Mahalanobis1936}
Mahalanobis, P. C. 1936,
Proc. Natl. Inst. Sci. India, 2, 49

\bibitem[Millon et al.(2020)]{Millon2020}
Millon, M., Galan, A., Courbin, F., et al. 2020,
A\&A, 639, A101

\bibitem[Nardini et al.(2019)]{Nardini2019}
Nardini, E., Lusso, E., Bisogni, S., et al. 2019,
A\&A, 632, A109

\bibitem[Pâris et al.(2018)]{Paris2018}
Pâris, I., Petitjean, P., Aubourg, É., et al. 2018,
A\&A, 613, A51

\bibitem[Pascale et al.(2025)]{Pascale2025}
Pascale, M., Frye, B. L., Pierel, J. D. R., et al. 2025,
ApJ, 979, 13

\bibitem[Pesce et al.(2020)]{Pesce2020}
Pesce, D. W., Braatz, J. A., Reid, M. J., et al. 2020,
ApJL, 891, L1

\bibitem[Planck Collaboration VI(2020)]{Planck2020}
Planck Collaboration VI. 2020,
A\&A, 641, A6

\bibitem[Ramos et al. (2020)]{Ramos2020}
Ramos et al. 2020,
MNRAS, 499, 3

\bibitem[Richards et al.(2001)]{Richards2001}
Richards, G. T., Fan, X., Schneider, D. P., et al. 2001,
AJ, 121, 2308

\bibitem[Riess et al.(2022)]{Riess2022}
Riess, A. G., Yuan, W., Macri, L. M., et al. 2022,
ApJL, 934, L7

\bibitem[Risaliti \& Lusso(2015)]{Risaliti2015}
Risaliti, G., \& Lusso, E. 2015,
ApJ, 815, 33

\bibitem[Risaliti \& Lusso(2019)]{Risaliti2019}
Risaliti, G., \& Lusso, E. 2019,
Nature Astronomy, 3, 272

\bibitem[Sacchi et al.(2022)]{Sacchi2022}
Sacchi, A., Risaliti, G., Signorini, M., et al. 2022,
A\&A, 663, L7

\bibitem[Schöneberg et al.(2022)]{Schoeneberg2022}
Schöneberg, N., Verde, L., Gil-Marín, H., \& Brieden, S. 2022,
JCAP, 11, 039

\bibitem[Scolnic et al.(2022)]{Scolnic2022}
Scolnic, D., Brout, D., Carr, A., et al. 2022,
ApJ, 938, 113

\bibitem[Scolnic et al.(2023)]{Scolnic2023}
Scolnic, D., Riess, A. G., Wu, J., et al. 2023,
ApJL, 954, L31

\bibitem[Shajib et al.(2023)]{Shajib2023}
Shajib, A. J., Mozumdar, P., Chen, G. C.-F., et al. 2023,
A\&A, 673, A9

\bibitem[Torres Arzayus et al.(2024)]{TorresArzayus2024}
Torres Arzayus, S., Delgado-Correal, C., Higuera-G., M. A., \& Rueda-Blanco, S. 2024,
Astrophys. Space Sci., 369, 17

\bibitem[Wang et al.(2023)]{Wang2023}
Wang, Y.-Y., Tang, S.-P., Jin, Z.-P., \& Fan, Y.-Z. 2023,
ApJ, 943, 13

\bibitem[York et al.(2000)]{York2000}
York, D. G., Adelman, J., Anderson, J. E., Jr., et al. 2000,
AJ, 120, 1579

\bibitem[Zajaček et al.(2024)]{Zajacek2024}
Zajaček, M., Czerny, B., Panda, S., et al. 2024,
Space Sci. Rev., 220, 29
\end{thebibliography}
\end{document}